\documentclass[useAMS,usenatbib, referee]{mn2e}
\usepackage{graphicx}
\usepackage{txfonts}

\def\apj{ApJ}
\def\apjl{ApJL}

\def\aap{A\&A}

\def\mnras{MNRAS}
\def\pasj{PASJ}

\def\beq#1{\begin{equation}\label{#1}}
\def\eeq{\end{equation}}
\def\beqa#1{\begin{eqnarray}\label{#1}}
\def\eeqa{\end{eqnarray}}

\def\Eq#1{Eq.~(\ref{#1})}

\def\myfrac#1#2{\left(\frac{#1}{#2}\right)}
\def\comment#1{\relax}

\newcommand{\ms}{M_\odot}
\newcommand{\msun}{M_\odot}

\title[Quasi-spherical accretion]{Theory of quasi-spherical accretion in X--ray
pulsars}
\author[N. Shakura et al.] {N. Shakura
\thanks{E-mail: nikolai.shakura@gmail.com, kpostnov@gmail.com},
K. Postnov, A. Kochetkova, 
L. Hjalmarsdotter\\
$^{1}$
Sternberg Astronomical Institute, Moscow State University, Universitetskij pr., 13, 119899, Moscow, Russia}

\begin{document}

\date{Received ... Accepted ...}
\pagerange{\pageref{firstpage}--\pageref{lastpage}} \pubyear{2010}

\maketitle

\label{firstpage}

\begin{abstract}
A theoretical model for quasi-spherical subsonic accretion onto slowly rotating magnetized 
neutron stars is constructed. In this model the accreting matter subsonically 
settles down onto the rotating magnetosphere forming an extended quasi-static shell.
This shell mediates the angular momentum removal from the rotating neutron star 
magnetosphere during spin-down episodes by large-scale convective motions. 
The accretion rate through the shell is determined by the ability of the plasma to enter the magnetosphere. 
The settling regime of accretion can be realized for moderate accretion rates 
$\dot M< \dot M_*\simeq 4\times 10^{16}$~g/s. At higher accretion rates a free-fall gap 
above the neutron star magnetosphere appears due to rapid Compton cooling, and accretion 
becomes highly non-stationary. 
From observations of the spin-up/spin-down rates (the angular rotation frequency derivative 
$\dot \omega^*$, and $\partial\dot\omega^*/\partial\dot M$ near the torque reversal) 
of X-ray pulsars with known orbital periods, it is possible 
to determine the main dimensionless parameters of the model, as well as to estimate the magnetic field of the neutron star.  
We illustrate the model by determining these parameters for three   
wind-fed X-ray pulsars GX 301-2, Vela X-1, and GX 1+4.   
The model explains both the spin-up/spin-down 
of the pulsar frequency on large time-scales and the irregular short-term frequency
fluctuations, which can correlate or anti-correlate with the X-ray flux fluctuations in 
different systems. It is shown that in real pulsars an almost iso-angular-momentum rotation law with $\omega \sim 1/R^2$, due to strongly anisotropic radial turbulent motions sustained by large-scale convection, is preferred. 
\end{abstract}


\section{Introduction}
\label{intro}
X-ray pulsars are highly magnetized neutron stars in binary systems, accreting matter from a companion star. The companion may be a low-mass star overfilling its Roche lobe in which case an accretion disc is formed. In the case of a high-mass companion, the neutron star may also accrete from the strong stellar wind and depending on the conditions a disc may be formed or accretion may take place quasi-spherically. The strong magnetic field of the neutron star disrupts the accretion flow at some distance from the neutron star surface and forces the accreted matter to funnel down on the polar caps of the neutron star creating hot spots that, if misaligned with the rotational axis, make the neutron star pulsate in X-rays. Most accreting pulsars show stochastic variations in their spin frequencies as well as in their luminosities. Many sources also exhibit long-term trends in their spin-behaviour with the period more or less steadily increasing or decreasing and in some sources spin-reversals have been observed. (For a thorough review, see e.g. Bildsten 1997 and references therein.)

The best-studied case of accretion is that of thin disc accretion (Shakura \& Sunyaev 1973). Here the spin-up/spin-down mechanisms are rather well understood.
For disc accretion the spin-up torque is determined by the specific angular momentum at the inner edge of the disc and can be written in the form 
$
K_{su}\approx \dot M\sqrt{GMR_A}\,
$
(Pringle and Rees, 1972). For a pulsar the inner radius of the accretion disc is determined by the Alfven radius $R_A$,  
$R_A\sim \dot M^{-2/7}$, so $K_{su}\sim \dot M^{6/7}$, i.e. 
for disc accretion the spin-up torque is weakly (almost linearly) dependent on the accretion rate (X-ray luminosity). 
In contrast, the spin-down torque for disc accretion in the first approximation is independent of $\dot M$: 
$K_{sd}\sim -\mu^2/R_c^3$, where 
$R_c=(GM/(\omega^*)^2)^{1/3}$ is the corotation radius, $\omega^*$ is the neutron star angular frequency and $\mu$ is
the neutron star's dipole magnetic moment. In fact, accretion torques in disc accretion are determined by a complicated disc-magnetospheric interaction, see, e.g., 
Ghosh \& Lamb (1979), Lovelace et al. (1995) and the 
discussion in Klu\'zniak \& Rappaport (2007), and correpsondingly can have a 
more complicated dependence on the mass accretion rate and other parameters.

Measurements of spin-up/spin-down in X-ray pulsars can be used to evaluate a very important parameter of the neutron star -- its magnetic field. The period of the pulsar is usually close to the equilibrium value $P_{eq}$, which is determined by the total zero torque applied to the neutron star, $K=K_{su}+K_{sd}=0$. So assuming the observed 
value $\omega^*=2\pi/P_{eq}$, the magnetic field of the neutron star in disc-accreting X-ray pulsars can be estimated if $\dot M$ is known. 
  
In the case of quasi-spherical accretion, which can take place in systems where the optical star underfills its Roche lobe and no accretion disc is formed,
the situation is more complicated. Clearly, the amount and sign of the 
angular mometum supplied to 
the neutron star from the captured stellar wind are important for spin-up or spin-down. To within a numerical factor (which can be positive or negative, see numerical simulations
by Fryxell \& Taam (1988), Ruffert (1997), Ruffert (1999), etc.), 
the torque applied to the neutron star 
in this case should be proportional to $\dot M \omega_B R_B^2$, where $\omega_B=2\pi/P_B$ is the binary orbital angular frequency, $R_B=2GM/(V_w^2+v_{orb}^2)^2$ 
is the gravitational capture (Bondi) radius, $V_w$ is the stellar wind velocity at the neutron star orbital distance, and $v_{orb}$ is the neutron star orbital velocity. In real high-mass X-ray binaries the orbital eccentricity is non-zero, the stellar wind 
is variable and can be inhomogeneous, etc., so $K_{su}$ can be a complicated function of time. The spin-down torque is
even more uncertain, since it is impossible to write down a simple equation like $-\mu^2/R_c^3$ any more
($R_c$ has no meaning for quasi-spherical accretion; for slowly rotating pulsars 
it is much larger than the Alfven radius where the angular momentum transfer from the accreting matter to the magnetosphere actually occurs). For example, if one uses
for the braking torque $-\mu^2/R_c^3$, the magnetic field 
in long-period X-ray pulsars turns out very high. We think this is a result
of underestimating the braking torque.

The matter captured from the stellar wind can accrete onto the neutron star 
in different ways. Indeed, if the X-ray flux from the accreting neutron star 
is sufficiently high, the shocked matter rapidly cools down due to Compton processes
and freely falls down toward the magnetosphere. The velocity of motion 
rapidly becomes supersonic, so a shock is formed above the magnetosphere. This regime was considered, e.g., by Burnard et al. (1983). Depending on the 
sign of the specific angular momentum of falling 
matter (prograde or retrograde), the neutron star can spin-up or spin-down. However, 
if the X-ray flux at the Bondi radius is below some value, 
the shocked matter remains hot, the radial velocity of the plasma is subsonic, and
the settling accretion regime sets in. A hot quasi-static shell forms around the magnetosphere (Davies \& Pringle, 1981). Due to additional energy release
(especially near the base of the shell), the temperature gradient across the shell
becomes superadiabatic, so large-scale convective motions inevitably appear. 
The convection intitiates turbulence, and the motion of a fluid element in the shell 
becomes quite complicated. If the magnetosphere allows plasma entry via instabilities
(and subsequent accretion onto the neutron star), the actual accretion rate
through such a shell is controlled by the magnetosphere (for example, the shell 
can exist, but the accretion through it can be weak or even absent altogether). 
So on top of the convective motions, the mean radial velocity of matter toward the magnetosphere, a subsonic settling, appears. This picture of accretion is realized 
at relatively small X-ray luminositites, $L_x<4\times 10^{36}$~erg/s (see below), 
and is totally different from 
what was considered in numerical simulations cited above.   
If the shell is present, its interaction with the rotating 
magnetosphere can lead to spin-up or spin-down of the neutron star, depending 
on the sign of the angular velocity difference between the 
accreting matter and the magnetospheric 
boundary. So in the settling accretion regime, both spin-up or spin-down 
of the neutron star is possible, even if the sign of the specific angular momentum of 
captured matter is prograde. The shell here mediates the angular momentum transfer
to or from the rotating neutron star.

One can find several models in the literature (see especially Illarionov \& Kompaneets 1990 and Bisnovatyi-Kogan 1991), from which the expression for the spin-down torque 
for quasi-spherically accreting neutron stars in the form 
$K_{sd}\sim -\dot M R_A^2 \omega^*\sim -\dot M^{3/7}$ can be derived.
Moreover, the expression for the Alfven radius 
$R_A$ in the case of settling accretion is found to have different dependence
on the mass accretion rate $\dot M$ and neutron star magnetic moment $\mu$, 
$\sim \dot M^{-2/11}\mu^{6/11}$, than the standard expression for disc
accretion, $\sim \dot M^{-2/7}\mu^{4/7}$, so 
the spin-down torque for quasi-spherical
settling accretion depends on the accretion rate as $K_{sd}\sim -\dot M^{3/11}$ (see below).
 
To stress the difference between quasi-spherical and disc accretion, it is
also instructive to rewrite the expression for the spin-down torque using the corotation and Alfven radii 
as $K_{sd}\sim -\mu^2/\sqrt{R_c^3 R_A^3}\sim -\mu^2/R_c^3(R_c/R_A)^{3/2}$ (see more detail below in Section 4). Since the factor $(R_c/R_A)^{3/2}\sim 
(\omega_K(R_A)/\omega^*)$ can be of the order of 10 or higher in real systems, 
using a braking torque in the form $\mu^2/R_c^3$ leads to a strong overestimation
of the magnetic field. 

The dependence of the braking torque on the accretion rate in the case 
of quasi-spherical settling accretion
suggests that variations of the mass accretion rate (and X-ray luminosity) 
must lead to a transition from spin-up (at high accretion rates) to spin-down (at small accretion rates) at some critical
value of $\dot M$ (or $R_A$), that differs from source to source.
 This phenomenon 
(also known as torque reversal) 
is actually observed in wind-fed pulsars like Vela X-1, GX 301-2 and GX 1+4,
which we shall consider below in more detail.

The structure of this paper is as follows. In Section 2, we present an outline of the theory for quasi-spherical accretion onto 
a neutron star magnetosphere. We show that it is possible to construct a hot envelope around the neutron star through which accretion can take place and act to either spin up or spin down the neutron star. In Section 3, we discuss the structure of the interchange instability region which determines whether the plasma can enter the magnetosphere of the rotating neutron star. In Section 4 we consider how the spin-up/spin-down torques vary with a changing accretion rate. In Section 5, we show how to determine the parameters of quasi-spherical accretion and the neutron star magnetic field from observational data. In Section 6, we apply our methods to the specific pulsars GX 301-2, Vela X-1 and GX 1+4. In Section 7 we discuss our results and, finally, in Section 8 we present our conclusions. A detailed gas-dynamic
treatment of the problem is presented in four appendices, which are very important to understand the physical processes involved. 

\section{Quasi-spherical accretion}
\label{s_qsaccr}

\subsection{The subsonic Bondi accretion shell}
\label{s_shell}
We shall here consider the torques applied to a neutron star in the case of quasi-spherical accretion from a stellar wind.
Wind matter is gravitationally captured by the moving neutron star and a bow-shock is formed at a characteristic 
distance $R\sim R_B$, where $R_B$ is the Bondi radius. Angular momentum can be removed from the neutron star 
magnetosphere in two ways --- either with matter expelled from the magnetospheric boundary without accretion (the propeller regime, Illarionov \& Sunyaev 1975), 
or via convective motions, which bring away angular momentum 
in a subsonic quasi-static shell around the magnetosphere, with accretion (the settling accretion regime).

In such a quasi-static shell, the temperature will be high (of the order of the virial temperature, see Davies and Pringle (1981)), and 
the important point is whether hot matter from the shell can in fact enter the magnetosphere. Two-dimensional calculations by Elsner and Lamb (1977) have shown that hot monoatomic ideal plasma is stable relative to the Rayleigh-Taylor instability at the magnetospheric boundary, and plasma cooling is thus needed for accretion to begin. However, a closer inspection of the 3-dimensional calculations by Arons and Lea (1976a) reveals that the hot plasma is only marginally stable at the magnetospheric equator (to within 5\% accuracy of their calculations). 
Compton cooling and the possible presence of dissipative phenomena (magnetic reconnection etc.) facilitates the plasma entering the magnetosphere. 
We will show that both accretion of matter from a hot envelope 
and spin-down of the neutron star is indeed possible.

\subsection{The structure of the shell around a neutron star magnetosphere}

To a zeroth approximation, we can neglect both rotation and radial motion (accretion) of matter
in the shell and consider only its hydrostatic structure.
The radial velocity of matter falling through the shell $u_R$ is   
lower than the sound velocity $c_s$. Under these assumptions,  
the characteristic cooling/heating time-scale 
is much larger than the free-fall time-scale. 

In the general case where both gas pressure and anisotropic turbulent motions are present, 
Pascal's law is violated.  Then the hydrostatic equilibrium equation can be derived 
from the equation of motion \Eq{v_R1} with stress tensor components \Eq{W_RR} - \Eq{W_pp}
and zero viscosity (see Appendix A for more detail): 
\beq{e1}
-\frac{1}{\rho}\frac{dP_g}{dR}-
\frac{1}{\rho R^2}\frac{d(P_\parallel^t R^2)}{dR}+\frac{2P_\perp^t}{\rho R}-\frac{GM}{R^2}=0
\eeq
Here $P_g=\rho c_s^2/\gamma$ is the gas pressure, and $P^t$ stands for the pressure due to turbulent 
motions:
\beq{ppar}
P_\parallel^t =\rho <u_\parallel^2>=\rho m_\parallel^2 c_s^2=\gamma P_g  m_\parallel^2
\eeq
\beq{pperp}
P_\perp^t =\rho <u_\perp^2>=\rho m_\perp^2 c_s^2 =\gamma P_g  m_\perp^2
\eeq
($<u_t^2>=<u_\parallel^2>+2<u_\perp^2>$ is the  
turbulent velocity dispersion, $m_\parallel^2$ and $m_\perp^2$ are 
turbulent Mach numbers squared in the radial and tangential directions, respectively;
for example, in the case of isotropic turbulence $m_\parallel^2=m_\perp^2=(1/3)m_t^2$ where $m_t$ is the turbulent Mach number).
The total pressure is the sum of the gas and turbulence terms: $P_g+P_t=P_g(1+\gamma m_t^2)$.

We shall consider, to a first approximation, that the entropy distribution in the shell is constant. 
Integrating the hydrostatic
equilibrium equation \Eq{e1}, we readily get 
\beq{hse_sol}
\frac{{\cal R} T}{\mu_m} = \myfrac{\gamma-1}{\gamma}\frac{GM}{R}\myfrac{1}
{1+\gamma m_\parallel^2-2(\gamma-1)(m_\parallel^2-m^2_\perp)}=\frac{\gamma-1}{\gamma}\frac{GM}{R}\psi(\gamma, m_t)\,.
\eeq
(In this solution we have neglected the integration constant, which is not important 
deep inside the shell. It is important in the outer part of the shell, but since the outer region close to the bow shock at $\sim R_B$ is not spherically 
symmetric, its structure can be found only numerically). 

Note that taking turbulence into account somewhat decreases the temperature within the shell.
Most important, however, is that the anisotropy of turbulent motions, caused
by convection in the stationary case, changes the distribution of the angular velocity 
in the shell. Below we will show that in the case of isotropic turbulence, the 
angular velocity distribution within the shell is close to the quasi-Keplerian one, 
$\omega(R) \sim R^{-3/2}$. In the case of strongly anisotropic turbulence caused by convection, $m_\parallel^2\gg m_\perp^2$, an approximately iso-angular-momentum distribution, $\omega(R) \sim R^{-2}$ is realized within the shell. Below we shall see that teh analysis 
of rela X-ray pulsars favors the iso-angular-momentum rotation distribution. 

Now, let us write down how the density varies inside the quasi-static shell for $R\ll R_B$.
For a fully ionized gas with $\gamma=5/3$ we find:
\beq{rho(R)}
\rho(R)=\rho(R_A)\myfrac{R_A}{R}^{3/2}
\eeq
and for the gas pressure:
\beq{P(R)}
P(R)=P(R_A) \myfrac{R_A}{R}^{5/2}\,.
\eeq
The above equations describe the structure of an ideal static adiabatic shell 
above the magnetosphere. Of course, at $R\sim R_B$ the problem is essentially non-spherically symmetric and
numerical simulations are required.  

Corrections to the adiabatic temperature gradient due to 
convective energy transport through the shell are calculated in Appendix C.

\subsection{The Alfven surface}

At the magnetospheric boundary (the Alfven surface), the total pressure (including
isotropic gas pressure and the possibly anisotropic turbulent pressure) is balanced by 
the magnetic pressure $B^2/(8\pi)$
\beq{}
P_g+P_t=P_g(R_A)(1+\gamma m_t^2)=\frac{B^2(R_A)}{8\pi}\,.
\eeq

The magnetic field at the Alfven radius is determined by 
the dipole magnetic field and by electric currents flowing on the Alfvenic surface
\beq{P(RA)}
P_g(R_A)=\frac{K_2}{(1+\gamma m_t^2)}\frac{B_0^2}{8\pi} \myfrac{R_0}{R_A}^6 =\frac{\rho{\cal R}T}{\mu_m}
\eeq
where the dimensionless coefficient $K_2$ takes into account the contribution from these currents 
and the factor $1/(1+\gamma m_t^2)$ is due to the turbulent pressure term.
For example, in the model by Arons and Lea (1976a, their Eq. 31), $K_2=(2.75)^2\approx 7.56$. 
At the magnetospheric cusp (where the magnetic force line is branched),
the radius of the Alfven surface is about 0.51 times that of the equatorial radius 
(Arons and Lea, 1976a). 
Below we shall assume that $R_A$ is the equatorial radius 
of the magnetosphere, unless stated otherwise. 

Due to the interchange instability, the plasma can enter the neutron star magnetosphere. 
In the stationary regime, let us introduce the accretion rate $\dot M$ onto the neutron star surface. 
From the continuity equation in the shell we find
\beq{rho_cont}
\rho(R_A)=\frac{\dot M}{4\pi u_R(R_A) R_A^2}
\eeq
Clearly, the velocity of absorption of matter by the magnetosphere is smaller than 
the free-fall velocity, so we introduce a dimensionless factor $f(u)=u_R/\sqrt{2GM/R}<1$.
Then the density at the magnetospheric boundary is
\beq{rho(R)}
\rho(R_A)=\frac{\dot M}{4\pi f(u) \sqrt{2GM/R_A} R_A^2}\,.
\eeq
For example, in the model calculations by Arons \& Lea (1976a) $f(u)\approx 0.1$;
in our case, at high X-ray luminosities the value of $f(u)$ can attain $\approx 0.5$.  
It is possible to imagine that the shell is impenetrable and that there is no accretion through it,  
$\dot M \to 0$. In this case $u_R\to 0$, $f(u)\to 0$, while the density 
in the shell remains finite. 
In some sense, the matter leaks from the magnetosphere down to the neutron star, 
and the leakage can be either small ($\dot M \to 0$) or large ($\dot M \ne 0$). 

Plugging $\rho(R)$ into \Eq{P(RA)} and using \Eq{hse_sol} and 
the definition of the dipole magnetic moment
\[
\mu=\frac{1}{2}B_0R_0^3
\] 
(where $R_0$ is the neutron star radius), we find
\beq{RA_def}
R_A=\left[\frac{4\gamma}{(\gamma-1)}\frac{f(u) K_2}{\psi(\gamma, m_t)(1+\gamma m_t^2)} \frac{\mu^2}{\dot M\sqrt{2GM}}\right]^{2/7}\,.
\eeq

It should be stressed that in the presence of a hot shell the Alfven radius is
determined by the static gas pressure at the magnetospheric boundary, 
which is non-zero even for a zero-mass accretion rate through the shell, 
so the appearance of $\dot M$ in the above formula is strictly formal. 

\subsection{Angular momentum transfer}

We now consider a quasi-stationary subsonic shell in which accretion proceeds 
onto the neutron star magnetosphere. We stress that in this regime, i.e. the settling regime,  the accretion rate onto the neutron star
is determined by the denisity at the bottom of the shell (which is directly related to 
the density downstream the bow shock in the gravitational capture region) and the ability of the plasma to enter the magnetosphere through the Alfven surface. 

The rotation law in the shell depends on the treatment of the turbulent viscosity
(see Appendix A for the Prandtl law and isotropic turbulence) and 
the possible anisotropy of the turbulence due to convection (see Appendix B). 
In the last case the anistropy leads to more powerful radial turbulence
than in the perpendicular directions. Thus, as shown in Appendix A and B,
there is a set of quasi-power-law solutions for the radial dependence of the angular 
rotation velocity in a convective shell. 
We shall consider a power-law dependence
of the angular velocity on radius, 
\beq{rotation_law}
\omega(R)\sim R^{-n}
\eeq
We will study in detail the quasi-Keplerian law with $n=3/2$, and the iso-angular-momentum distribution
with $n=2$, which in some sense are limiting cases
among the possible solutions. 

When approaching the bow shock, $R \to  R_B$,  
$\omega\to  \omega_B$. 
Near the bow shock the problem is not spherically
symmetric any more since the flow is more complicated (part of the flow bends across the shell),
and the structure of the flow can be studied only using numerical simulations.
In the absence of such simulations, we shall assume that the 
iso-angular-momentum distribution is valid up to the nearest distance to 
the bow shock from the neutron star which we shall take to be the gravitational
capture radius $R_B$, 
\[
R_B\simeq 2GM/(V_w^2+v_{orb}^2)^2
\]
where $V_w$ is the stellar wind velocity at the neutron star orbital distance, and $v_{orb}$ is the neutron star orbital velocity.

This means that the angular velocity of rotation of matter near the magnetosphere $\omega_m$
will be related to $\omega_B$ via
\beq{omega_m1}
\omega_m= \tilde\omega\omega_B\myfrac{R_B}{R_A}^{n}.
\eeq
(Here the numerical factor $\tilde\omega>1$ takes into account the deviation 
of the actual rotational law from the value obtained by using the assumed power-law dependence near the Alfven surface; see Appendix A for more detail.)

Let the NS magnetosphere rotate with 
an angular velocity $\omega^*=2\pi/P^*$ where $P^*$ is the neutron star spin period. 
The matter at the bottom of the shell 
rotates with an angular velocity $\omega_m$, in general different from $\omega^*$.
If $\omega^*>\omega_m$, coupling of the plasma with the magnetosphere ensures transfer of angular momentum from the magnetosphere to the shell, or from the shell to the
magnetosphere if $\omega^*<\omega_m$. In the general case, the coupling of matter with the magnetosphere can be moderate or 
strong. In the strong coupling regime the toroidal magnetic field component $B_t$ is proportional to the 
poloidal field component $B_p$ as $B_t\sim -B_p (\omega_m-\omega^*)t$, and $|B_t|$ can grow to $\sim |B_p|$. This regime can be expected for rapidly rotating magnetopsheres 
when $\omega^*$ is comparable to or even greater than the Keplerian
angular frequency $\omega_K(R_A)$; in the latter case the propeller regime sets in. 
In the moderate coupling regime, the plasma can enter the magnetosphere due to instabilities on a timescale shorter than that
needed for the toroidal field to grow to the value of the poloidal field, 
so $B_t < B_p$.  

\subsubsection{The case of strong coupling}
\label{s:strongcoupling}

Let us first consider the strong coupling regime. In this regime, powerful large-scale convective motions can lead to turbulent magnetic field diffusion 
accompanied by magnetic field dissipation. This process is characterized by 
the turbulent magnetic field diffusion coefficient $\eta_t$.  In this case 
the toroidal magnetic field (see e.g. Lovelace et al. 1995 
and references therein) is
\beq{bt}
B_t=\frac{R^2}{\eta_t}(\omega_m-\omega^*)B_p\,.
\eeq
The turbulent magnetic diffusion coefficient is related to the kinematic turbulent viscosity as
$\eta_t\simeq \nu_t$. The latter can be written as
\beq{nut}
\nu_t=<u_tl_t>\,.
\eeq 
According to the phenomenological Prandtl law which relates the average characteristics of 
a turbulent flow (the velocity $u_t$, the characteristic scale of turbulence $l_t$ 
and the shear $\omega_m-\omega^*$)
\beq{Prandtl}
u_t\simeq l_t |\omega_m-\omega^*|\,.
\eeq
In our case, the turbulent scale must be 
determined by the largest scale of the energy supply to turbulence 
from the rotation of a non-spherical magnetospheric surface. This scale is determined by the velocity difference of the
solidly rotating magnetosphere and the accreting matter that is still not interacting with the magnetosphere, i.e. $l_t\simeq R_A$, which determines the turn-over velocity of the largest turbulence eddies. At smaller scales a turbulent cascade develops. 
Substituting this scale into equations \Eq{bt}-\Eq{Prandtl} above, we   
 find that in the strong coupling regime $B_t\simeq B_p$. The 
moment of forces due to plasma-magnetosphere interactions is applied to the neutron star and 
causes spin evolution according to: 
\beq{}
I\dot \omega^*=\int\frac{B_tB_p}{4\pi}\varpi dS = 
\pm \tilde K(\theta)K_2\frac{\mu^2}{R_A^3}
\eeq
where $I$ is the neutron star's moment of inertia, $\varpi$ is the distance from the rotational axis and $\tilde K(\theta)$ is a numerical coefficient depending
on the angle between the rotational and magnetic dipole axis. The coefficient
$K_2$ appears in the above expression for the same reason as in \Eq{P(RA)}.
The positive sign corresponds to positive flux of angular momentum to the neutron star 
($\omega_m>\omega^*$). The negative sign corresponds to negative flux of angular momentum across 
the magnetosphere ($\omega_m<\omega^*$). 

At the Alfven radius, the matter couples with the magnetosphere and acquires the angular velocity of the neutron star. It then falls onto the neutron-star surface
and returns the angular momentum acquired at $R_A$ back to the neutron star via the magnetic field. 
As a result of this process, the neutron star spins up at a rate determined by the expression:
\beq{suz}
I\dot \omega^*=+z \dot M R_A^2\omega^*
\eeq
where $z$ is a numerical coefficient which takes into account the angular momentum of the falling matter. If all matter falls from the equatorial equator, $z=1$; if matter falls strictly along the spin axis, $z=0$. 
If all matter were to fall across the entire magnetospheric surface, then $z=2/3$. 

Ultimately, the total torque applied to the neutron star in the strong coupling regime yields
\beq{sd_eq_strong}
I\dot \omega^*=\pm \tilde K(\theta)K_2 \frac{\mu^2}{R_A^3} +z \dot M R_A^2\omega^*
\eeq

Using \Eq{RA_def}, we can eliminate $\dot M$ in the above equation to obtain in the spin-up regime ($\omega_m>\omega^*$)
\beq{su}
I\dot \omega^*=\frac{\tilde K(\theta)K_2\mu^2}{R_A^3}
\left[1+z\frac{4\gamma f(u)}{\sqrt{2}(\gamma-1)(1+\gamma m_t^2)\psi(\gamma, m_t)\tilde K(\theta)}\myfrac{R_A}{R_c}^{3/2}\right]
\eeq
where $R_c^3=GM/(\omega^*)^2$ is the corotation radius. In the spin-down regime ($\omega_m<\omega^*$) we find
\beq{sd}
I\dot \omega^*=-\frac{\tilde K(\theta)K_2\mu^2}{R_A^3}
\left[1-z\frac{4\gamma f(u)}{\sqrt{2}(\gamma-1)(1+\gamma m_t^2)\psi(\gamma, m_t)\tilde K(\theta)}\myfrac{R_A}{R_c}^{3/2}\right]\,.
\eeq
Note that in both cases $R_A$ must be smaller than $R_c$, otherwise the propeller effect prohibits accretion. In the propeller regime $R_A>R_c$, 
matter does not fall onto the neutron star, there are no accretion-generated 
X-rays from the neutron star, the shell rapidly cools down and
shrinks and the standard Illarionov and Sunyaev propeller (1975), with matter outflow from the magnetosphere is established.

In both accretion regimes (spin-up and spin-down), 
the neutron star angular velocity $\omega^*$ almost approaches the angular velocity of
matter at the magnetospheric boundary, $\omega^*\to \omega_m(R_A)$. The difference between $\omega^*$ and
$\omega_m$ is small when the second term in the square brackets in \Eq{su} and \Eq{sd} is much smaller than unity. Also note that when approaching the propeller regime ($R_A\to R_c$), the accretion rate decreases, 
$f(u)\to 0$, the second term in the square brackets vanishes, and the spin evolution is determined solely by the spin-down term $-\tilde K(\theta) \mu^2/R_A^3$.
(In the
propeller regime, $\omega_m< \omega_K(R_A)$, $\omega_m<\omega^*$, $\omega^*> \omega_K(R_A)$ ). So the neutron star spins down to the Keplerian frequency at the Alfven radius.
In this regime, the specific angular momentum of the matter that flows in and out
from the magnetosphere is, of course, conserved. 

Near the equilibrium accretion state ($\omega^*\sim \omega_m$),
relatively small fluctuations in $\dot M$ across the shell would lead to very strong fluctuations in 
$\dot \omega^*$ since the toroidal field component can change its sign
by changing from $+B_p$ to $-B_p$. This property, if realized in nature, could be 
the distinctive feature of the strong coupling regime. It is known (see eg.g. Bildsten et al. 1997, Finger et al. 2011) that 
real X-ray pulsars sometimes exhibit rapid spin-up/spin-down transitions not associated
with X-ray luminosity changes, which may be evidence for 
them temporarily entering the strong coupling regime. It is not excluded that the
triggering of the strong coupling regime may be due to the magnetic field
frozen into the accreting plasma that has not yet entered the magnetosphere. 


\subsubsection{The case of moderate coupling}

The strong coupling regime considered above can be realized in the extreme case where 
the toroidal magnetic field $B_t$ attains a maximum possible value  
$\sim B_p$ due to magnetic turbulent diffusion. 
Usually, the coupling of matter with the magnetosphere is mediated by 
different plasma instabilities whose characteristic times are too short for substantial toroidal field growth. 
As we discussed above in Section \ref{s_shell}, the shell is very hot near the
bottom, so without cooling at the magnetospheric boundary it is marginally stable with respect to the interchange instability, according to the calculations by Arons 
and Lea (1976a). Due to Compton cooling by X-ray emission from the neutron star poles, plasma enters the magnetosphere
with a characteristic time-scale determined by the instability increment. Since the most likely instability is the Rayleigh-Taylor instability, 
the growth rate scales as the Keplerian angular frequency $\omega_K=\sqrt{GM/R^3}$. 
The time-scale of the instability can be normalized as  
\beq{}
t_{inst}=\frac{1}{\omega_K(R_A)}\,,
\eeq
The toroidal magnetic field increases with time as 
\beq{BtBp}
B_t=K_1(\theta) B_p(\omega_m-\omega^*)t_{inst}
\eeq 
where $K_1(\theta)$ is a numerical coefficient which takes into account the degree and angular dependence of the coupling of matter with the magnetosphere in which the angle between the neutron star spin axis and the magnetic dipole axis is $\theta$. 
Then, the 
neutron star spin frequency change in this regime reads
\beq{sd1}
I\dot\omega^*=K_1(\theta)K_2\frac{\mu^2}{R_A^3}\frac{\omega_m-\omega^*}{\omega_K(R_A)}\,.
\eeq

Using the definition of $R_A$ [\Eq{RA_def}] and $\omega_K$, the spin-down formula can be recast to the form
\beq{sd_om}
I\dot \omega^*=Z \dot M R_A^2(\omega_m-\omega^*)
\eeq
Here the dimensionless coefficient $Z$ is 
\beq{Zdef}
Z=\frac{ K_1(\theta)}{f(u)}\frac{\sqrt{2}(\gamma-1)}{4\gamma}\psi(\gamma, m_t)
(1+\gamma m_t^2)\,.
\eeq

Taking into account that the matter falling onto the neutron star surface brings angular momentum $z\dot M R_A^2\omega^*$
(see \Eq{suz} above), we arrive at 
\beq{sd_eq}
I\dot \omega^*=Z \dot M R_A^2(\omega_m-\omega^*)+z \dot M R_A^2\omega^*
\eeq
Clearly, to remove angular momentum from the neutron star through this kind of a static shell, $Z$ must be larger than $z$. Then the neutron star can spin-down episodically (we shall precise this statement below). Oppositely, if $Z<z$, the neutron star can only spin up. 

When a hot shell cannot be formed (at high accretion rates or small relative wind velocities, see e.g. Sunyaev (1978)), free-fall Bondi accretion with low angular momentum is realized. No angular momentum can be removed from the neutron star magnetosphere. 
Then $Z=z$ and \Eq{sd_eq} takes the simple form $
I\dot \omega^*=Z \dot M R_A^2\omega_m $, and the neutron star in this regime can spin-up up to about $\omega_K(R_A)$ independent of the sign of the difference of the angular frequencies 
$\omega_m-\omega^*$ at the magnetopsheric boundary.   
Due to conservation of specific angular momentum, $\omega_m=\omega_B (R_B/R_A)^2$, so 
in this case the spin evolution of the NS is described by equation
\beq{}
I\dot \omega^*=Z \dot M \omega_BR_B^2\,,
\eeq 
where $Z$ plays the role of the specific angular momentum
of the captured matter. For example, in early work by Illarionov \& Sunyaev 1975
$Z\simeq 1/4$. However, detailed numerical simulations of Bondi-Littleton accretion 
in 2D (e.g. Fryxell and Taam, 1988, Ho et al. 1989) and in 3D (e.g. Ruffert, 1997, 1999) 
revealed that due to inhomogeneities in the incoming flow, a non-stationary 
regime with an alternating sign of the captured matter angular momentum can be realized. So
the sign of $Z$ can be negative as well, and alternating spin-up/spin-down regimes
can be observed. Such a scenario is frequently invoked to explain the observed torque
reversals in X-ray pulsars (see the discussion in Nelson et al. 1997). We repeat that 
this could indeed be the case for large X-ray luminosities $>4\times 10^{36}$~erg/s when
a convective quasi-hydrostatic shell cannot exist due to strong Compton cooling 
near the magnetospheric boundary.

When a hot shell is formed (at moderate X-ray luminositities 
below $\sim 4\times 10^{36}$~erg/s, see \Eq{M*} below), 
the angular momentum from the neutron star
magnetosphere can be transferred away through the shell by turbulent viscosity and 
convective motions. So we substitute 
$\omega_m$ from \Eq{omega_m1} into \Eq{sd_eq}
to obtain
\beq{sd_eq1}
I\dot \omega^*= Z\dot M \tilde\omega\omega_B R_B^2\myfrac{R_A}{R_B}^{2-n}-Z(1-z/Z)\dot M R_A^2\omega^*\,.
\eeq
This is the main formula which we shall use below. 
To proceed further, however, we need to determine the dimensionless coefficients of
this equation. In the next section we shall find the important factor $f(u)$ that 
enters the formulae for both $Z$ and $R_A$, so the only unknown dimensionless parameter 
of the problem will be the coefficient $K_1(\theta)$. 

\section{Approximate structure of the interchange instability region}

The plasma enters the magnetosphere of the slowly rotating neutron star due to 
the interchange instability. The boundary between the plasma and the magnetosphere is stable 
at high temperatures $T>T_{cr}$, but becomes unstable at $T<T_{cr}$, and remains in 
a neutral equilibrium at $T=T_{cr}$ (Elsner and Lamb, 1976). The critical temperature is:
\beq{Tcr}
{\cal R}T_{cr}=\frac{1}{2(1+\gamma m_t^2)}\frac{\cos\chi}{\kappa R_A}\frac{\mu_mGM}{R_A}
\eeq
Here $\kappa$ is the local curvature of the magnetosphere, $\chi$ is the angle 
the outer normal makes with the radius-vector at a given point, and the contribution 
of turbulent pulsations in the plasma to the total pressure is
taken into account by factor $(1+\gamma m_t^2)$.
The effective gravity acceleration can be written as 
\beq{g_eff}
g_{eff}=\frac{GM}{R_A^2}\left(1-\frac{T}{T_{cr}}\right)\,.
\eeq
The temperature in the quasi-static shell is given by \Eq{hse_sol}, so 
the condition for the magnetosphere instability can then be rewritten as:
\beq{m_inst}
\frac{T}{T_{cr}}=\frac{2(\gamma-1)(1+\gamma m_t^2)}
{\gamma}\psi(\gamma, m_t) \frac{\kappa R_A}{\cos\chi}<1\,.
\eeq
According to Arons and Lea (1976a), when the external gas pressure decreases 
with radius as $P\sim R^{-5/2}$, the form of the magnetosphere far from the 
polar cusp can be described to within 10\% accuracy as $(\cos\lambda)^{0.2693}$
(here $\lambda$ is the polar angle counting from the magnetospheric equator). The instability first appears near the equator, where the curvature is minimal.
Near the equatorial plane ($\lambda=0$),
for a poloidal dependence of the magnetosphere $\approx (\cos \lambda)^{0.27}$ 
we get for the curvature $k_pR_A=1+1.27$. The toroidal field curvature at the magnetospheric equator is  $k_t=1$. The tangent sphere at the equator cannot have a radius larger than the inverse poloidal curvature, therefrom $\kappa R_A=1.27$ at $\lambda=0$.
This is somewhat larger than the value of $\kappa R_A=\gamma/(2(\gamma-1))=5/4=1.25$ (
for $\gamma=5/3$ in the absence of turbulence or
for fully isotropic turbulence),
but within the accuracy limit\footnote{In Arons and Lea (1976b), the curvature is calculated to be $\kappa R_A\approx 1.34$, still within the accuracy limit}.
The contribution from anisotropic turbulence 
decreases the critical temperature; for example,
for $\gamma=5/3$, in the case of strongly anisotropic turbulence 
$m_\parallel=1$, $m_\perp=0$, at $\lambda=0$ we obtain $T/T_{cr}\sim 2$, 
i.e. anisotropic turbulence
increases the stability of the magnetosphere.
So initially the plasma-magnetospheric boundary
is stable, and after cooling to $T<T_{cr}$ the plasma instability sets in, starting in the equatorial zone, where the curvature of the magnetospheric surface is minimal.   

Let us consider the development of the interchange instability when cooling 
(predominantly the Compton cooling) is present. The temperature changes as (Kompaneets, 1956, 
Weymann, 1965)
\beq{dTdt}
\frac{dT}{dt}=-\frac{T-T_x}{t_C}
\eeq 
where the Compton cooling time is
\beq{t_comp}
t_{C}=\frac{3}{2\mu_m}\frac{\pi R_A^2 m_e c^2}{\sigma_T L_x}
\approx 10.6  [\hbox{s}] R_{9}^2 \dot M_{16}^{-1}\,.
\eeq
Here $m_e$ is the electron mass, $\sigma_T$ is the Thomson cross section, $L_x=0.1 \dot M c^2$ is the X-ray luminosity, $T$ is the electron temperature (which is equal to ion temperature; the timescale of electron-ion energy exchange is the shortest one), $T_x$ is the X-ray temperature and
$\mu_m=0.6$ is the molecular weight.  
The photon temperature is $T_x=(1/4) T_{cut}$ for a bremsstrahlung spectrum with an exponential 
cut-off at $T_{cut}$, typically $T_x=3-5$~keV. 
The solution of equation \Eq{dTdt} reads:
\beq{}
T=T_x+(T_{cr}-T_x)e^{-t/t_C}\,.
\eeq
It is seen that for $t\approx 2t_C$ the temperature decreases to $T_x$. 
In the linear approximation and noticing that $T_{cr}\sim 30\,\hbox{keV}\gg T_x\sim 3$~keV, 
the effective gravity acceleration increases linearly with time:
\beq{}
g_{eff}\approx \frac{GM}{R_A^2}\frac{t}{t_C}\,.
\eeq
Correspondingly, the rate of instability increases with time as
\beq{}
u_{i}=\int g_{eff} dt=\frac{1}{2}\frac{GM}{R_A^2}t^2\,.
\eeq
Let us introduce the mean rate of the instability growth
\beq{}
<u_i>=\frac{\int u dt}{t}=\frac{1}{6}\frac{GM}{R_A^2}\frac{t^2}{t_C}=
\frac{1}{6}\frac{GM}{R_A^2t_C}\myfrac{\zeta R_A}{<u_i>}^2\,.
\eeq
Here $\zeta\lesssim 1$ and $\zeta R_A$ is the characteristic scale of the
instability that grows with the rate $<u_i>$.
So for the mean rate of the instability growth in the linear stage we find
\beq{ui}
<u_i>=\myfrac{\zeta^2GM}{6t_C}^{1/3}=\frac{\zeta^{2/3}}{12^{1/3}}\sqrt{\frac{2GM}{R_A}}
\myfrac{t_{ff}}{t_C}^{1/3}\,.
\eeq
Here we have introduced the free-fall time as
\beq{}
t_{ff}=\frac{R_A^{3/2}}{\sqrt{2GM}}\,.
\eeq

Clearly, later in the non-linear stage the rate of instability growth approaches the
free-fall velocity. We consider the linear stage first of all, since 
at this stage the temperature is not too low (although the entropy 
starts decreasing with radius), and it is in this zone that the effective angular 
momentum transfer from the magnetosphere to the shell occurs. At later stages of the
instability development, 
the entropy drop is too strong for convection to begin. 

Let us estimate the accuracy of our approximation by retaining the second-order terms
in the exponent expansion. Then the mean instability growth rate is 
\beq{2dorder}
<u_i>=\myfrac{\zeta^2GM}{6t_C}^{1/3}\left[1-2\zeta^{1/3}\myfrac{t_{ff}}{t_C}^{2/3}\right]\,.
\eeq
Clearly, the smaller accretion rate, the smaller the ratio $t_{ff}/t_C$, and the better our approximation. 

Now we are in a position to specify the important dimensionless 
factor $f(u)$:
\beq{fu1}
f(u)=\frac{<u_i>}{u_{ff}(R_A)}
\eeq
Substituting \Eq{ui} and \Eq{fu1} into \Eq{RA_def}, we find for the Alfven radius in 
this regime:
\beq{RA}
R_A\approx 0.9\times 10^9[\hbox{cm}]
\left(\frac{4\gamma\zeta}{(\gamma-1)(1+\gamma m_t^2)\psi(\gamma, m_t)}\frac{\mu_{30}^3}{\dot M_{16}}\right)^{2/11}\,.
\eeq
We stress the difference of the obtained expression for the Alfven radius with the 
standard one, $R_A\sim \mu^{4/7}/\dot M^{-2/7}$, which is obtained by equating the dynamical
pressure of falling gas to the magnetic field pressure; this difference comes from the dependence of 
$f(u)$ on the magnetic moment and mass accretion rate in the settling accretion regime. 

Plugging \Eq{RA} into \Eq{fu1}, we obtain an explicit expression for $f(u)$:
\beq{fu}
f(u)\approx 0.33\myfrac{(\gamma-1)(1+\gamma m_t^2)\psi(\gamma, m_t)}{4\gamma\zeta}^{1/33}\dot M_{16}^{4/11}\mu_{30}^{-1/11}\,.
\eeq

\section{Spin-up/spin-down transitions}

Now let us consider how the spin-up/spin-down torques vary with changing $\dot M$.
We stress again that we consider the shell through which matter accretes to be essentially subsonic. 
It is the leakage of matter through the magnetospheric boundary 
that in fact determines the actual accretion rate onto the neutron star. This is 
mostly dependent on the density at the bottom of the shell. On the other hand, the density structure of
the shell is directly related to the density of captured matter at the bow shock region, 
so density variations downstream the shock are rapidly translated into density variations near the magnetopsheric boundary.  This means that the actual accretion rate variations must be  
essentially independent (for circular or low-eccentricity orbits) of the orbital phase
but are mostly dependent on variations of the wind density. In contrast, 
possible changes in $R_B$ (for example, due to wind velocity variations or to the orbital motion of the neutron star) 
do not affect the accretion rate through the shell but 
strongly affect the value of the spin-up torque (see \Eq{sd_eq1}).

\Eq{sd_eq1} can be rewritten in the form
\beq{sd_eq2}
I\dot \omega^*=A\dot M^{\frac{3+2n}{11}} - B\dot M^{3/11}\,.
\eeq
For the fiducial value $\dot M_{16}\equiv \dot M/10^{16}$~g/s, 
the accretion-rate independent coefficients are (in CGS units)
\beq{A(Z)}
A\approx 5.325\times 10^{31} (0.034)^{2-n}K_1(\theta)\tilde \omega 
\delta^n (1+(5/3) m_t^2)\myfrac{\zeta}{(1+(5/3) m_t^2)\psi(5/3,m_t)}^{\frac{13-6n}{33}}
\mu_{30}^{\frac{13-6n}{11}}v_8^{-2n}\myfrac{P_b}{10\hbox{d}}^{-1}
\eeq
\beq{B}
B=5.4\times 10^{32}(1-z/Z)K_1(\theta)\myfrac{\zeta}{(1+(5/3) m_t^2)\psi(5/3,m_t)}^{\frac{13}{33}}\mu_{30}^{\frac{13}{11}}\myfrac{P^*}{100\hbox{s}}^{-1}
\eeq
(from now on in numerical estimates we assume $\gamma=5/3$).
The dimensionless factor $\delta<1$ takes into account the actual location
of the gravitaional capture radius, which is smaller than the Bondi value for a cold stellar wind (Hunt, 1971). This radius can also be smaller due to 
radiative heating of the stellar wind by X-ray emission from the neutron star surface (see below).
In the numerical coefficients we have used the expression for $Z$ with account for \Eq{fu},
and \Eq{RA} for the Alfvenic radius. 

Below we shall consider the case $Z-z>0$, i.e. $B>0$, otherwize only spin-up of the neutron star is possible. 

\begin{figure*}
\includegraphics[width=0.5\textwidth]{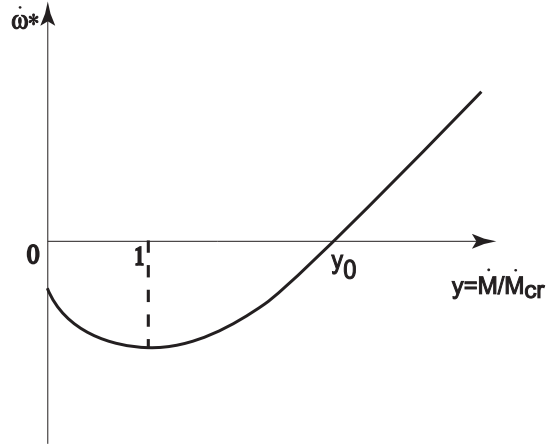}
\caption{A schematic plot of $\dot\omega^*$ as a function of $y$ [\Eq{sdy}]. In fact, as $y\to 0$, $\omega^*$ approaches some negative $\dot\omega^*$, since the neutron star 
enters the propeller regime at small accretion rates.}
\label{f:y}
 \end{figure*}

First of all, we note that the function $\dot\omega^*(\dot M$) reaches minimum at 
some $\dot M_{cr}$, By differentiating \Eq{sd_eq2} with respect to $\dot M$ and equating to zero, we find
\beq{dotMcr}
\dot M_{cr}=\left[\frac{B}{A}\frac{3}{(3+2n)}\right]^{\frac{11}{2n}}
\eeq 
At $\dot M=\dot M_{cr}$ the value of $\dot\omega^*$ reaches an absolute minimum (see Fig. \ref{f:y}).  

It is convenient to introduce the dimensionless parameter
\beq{y_def}
y\equiv \frac{\dot M}{\dot M_{cr}}
\eeq
and rewrite \Eq{sd_eq2} in the form
\beq{sdy}
I\dot \omega^*=A\dot M_{cr}^{\frac{3+2n}{11}}y^{\frac{3+2n}{11}}
\left(1- {\bf\myfrac{y_0}{y}^\frac{2n}{11}}\right)\,,
\eeq
where the frequency derivative vanishes at $y=y_0$:
\beq{y0}
y_0=\myfrac{3+2n}{3}^{\frac{11}{2n}}\,.
\eeq
The qualitative behaviour of $\dot\omega^*$ as a function of $y$ is shown in Fig. 
\ref{f:y}.

Let us then vary \Eq{sdy} with respect to $y$:
\beq{variations}
I(\delta\dot\omega^*)=I\frac{\partial \dot \omega^*}{\partial y}(\delta y)=
\frac{3+2n}{11}A\dot M_{cr}^\frac{3+2n}{11}y^{-\frac{8-2n}{11}}
\left(1- \frac{1}{y^{\frac{2n-1}{11}}}\right)(\delta y)\,.
\eeq
We see that depending on whether $y>1$ or $y<1$, 
\textit{correlated changes} of $\delta \dot\omega^*$
with X-ray flux should have different signs. Indeed, for GX 1+4 in Gonz\'alez-Gal\'an et al (2011) a positive correlation
of the observed $\delta P$ with $\delta \dot M$ was found using \textit{Fermi} data. This means that 
there is a negative correlation between $\delta\omega^*$ and $\delta\dot M$, suggesting $y<1$ in this source.

\section{Determination of the neutron star magnetic field and other parameters
in the settling accretion regime}
\label{s:magfield}

Most X-ray pulsars rotate close to their equilibrium periods, i.e. the 
average $\dot\omega^*=0$. Near the equilibrium, in the settling accretion regime 
from \Eq{sd_eq2} we obtain:
\beq{mu_eq}
\mu_{30}^{(eq)}\approx
\left[\frac{0.0986\cdot (0.034)^{(2-n)}\tilde\omega (1+(5/3)m_t^2)}{1-z/Z}\right]^\frac{11}{6n}
\myfrac{P_*/100s}{P_b/10d}^\frac{11}{6n}\myfrac{\dot M_{16}(1+(5/3) m_t^2)\psi(5/3,m_t)}{\zeta}^\frac{1}{3}
\myfrac{\sqrt{\delta}}{v_8}^\frac{11}{3}
\eeq

Once the magnetic field of the neutron star is estimated for any specific system, we can 
calculate the value of the Alfven radius $R_A$ [\Eq{RA}] and the 
important numerical coefficient $f(u)$ [\Eq{fu}]. The coupling constant $K_1(\theta)$
is evaluated from \Eq{A(Z)}, in which the left-hand side can be independently 
calculated using \Eq{variations} measured at $y=y_0$ (where $\dot \omega^*=0$):
\beq{Adet}
A=\frac{I\left.\myfrac{\partial \dot \omega^*}{\partial y}\right|_{y_0}}
{\myfrac{3+2n}{11}\frac{\dot M^\frac{3+2n}{11}}{y_0}
\left(1-y_0^{\frac{-(2n-1)}{11}}\right)}\,.
\eeq
The coefficient $Z$ is then determined from \Eq{Zdef}. The 
dimensionless factor relating the toroidal and poloidal magnetic field is also important. 
Near the equilibrium we have $\omega_m-\omega^*=-(z/Z)\omega^*$, so  
\Eq{BtBp} can be written as 
\beq{BtBpnum}
\frac{B_t}{B_p}=K_1(\theta)\myfrac{z}{Z}\myfrac{\omega^*}{\omega_K(R_A)}=
\frac{10f(u)z}{\sqrt{2}(1+(5/3)m_t^2)\psi(5/3,m_t)}\myfrac{\omega^*}{\omega_K(R_A)}\,.
\eeq

Calculated values for all parameters and coefficients discussed above are listed for 
specific wind-fed pulsars in Table 1 below. 

\subsection{Low-luminosity X-ray pulsars with torque reversal}

Let us consider X-ray pulsars with persistent spin-up and spin-down episodes. We shall assume that a convective shell is present during both spin-up (with higher luminosity) and spin-down (with lower luminosity), provided that at the spin-up stage the mass accretion rate
does not exceed $\dot M_*$ as derived above. 

Suppose that in such pulsars we can measure the average value of $\dot \omega^*|_{su}$ and $\dot \omega^*|_{sd}$ as well as the average X-ray
flux during spin-up and spin-down, respectively.  

Let at some specific value $y_1<y_0$ the source be observed to spin-down:
\beq{suy1} 
I\dot \omega^*|_{sd}=A\dot M_{cr}^{\frac{3+2n}{11}}y_1^{\frac{3+2n}{11}}
\left(1- \myfrac{y_0}{y_1}^{2n/11}\right)>0\,, \quad y_1<y_0=\myfrac{3+2n}{3}^{\frac{11}{2n}}\,.
\eeq
At some $y_2>y_0$, the neutron star starts to spin-up:
\beq{suy2} 
I\dot \omega^*|_{su}=A\dot M_{cr}^{\frac{3+2n}{11}}y_2^{\frac{3+2n}{11}}
\left(1- \myfrac{y_0}{y_2}^{2n/11}\right)<0\,, \quad y_2>y_0
\eeq
Using the observed spin-down/spin-up ratio $\dot\omega^*|_{sd}/\dot\omega^*|_{su}=X$ 
and the corresponding X-ray luminosity ratio $L_x(sd)/L_x(su)=y_1/y_2=x<1$, we find by dividing \Eq{suy1} with \Eq{suy2} and after substituting $y_2=y_1/x$
\beq{}
|X|=x^{\frac{3+2n}{11}}
\left|
\frac{1-\myfrac{y_0}{y_1}^{2n/11}}
{1-\myfrac{y_0}{(y_1/x)}^{2n/11}}
\right|\,.
\eeq
Solving this equation, we obtain  $y_1$ and $y_2$
through the observed quantities $|X|$ and $x$. We stress that so far we have not used the absolute values of
the X-ray luminosity (the mass accretion rate), only the ratio of the X-ray fluxes during spin-up and spin-down. 

Here we should emphasize an important point. When the accretion rate through the shell 
exceeds some critical value $\dot M>\dot M^*$, the flow near the Alfven surface may 
become supersonic, a free-fall gap appears above the magnetosphere, and the 
angular momentum can not be removed from the magnetosphere. In that case the 
settling accretion regime is no longer realized, a shock is formed in the flow near
the magnetosphere (the case studied by Burnard et al. 1982). Depending on the
character of the inhomogeneities in the captured stellar wind, 
the specific angular momentum of the accreting matter can be prograde or retrograde, 
so alternating spin-up and spin-down episodes  
are possible. Thus the transition from the settling accretion regime 
(at low X-ray luminosities) to Bondi-Littleton accretion 
(at higher X-ray luminosities) can 
actually occur before $y$ reaches $y_0$. Indeed, by assuming a maximum 
possible value of the dimensioless velocity of matter settling $f(u)$=0.5 (for angular momentum removal from the magnetosphere to be possible, 
see Appendix D for more detail), we find from \Eq{fu} :
\beq{M*}
\dot M^*_{16}\approx 3.7 \myfrac{\zeta}{(1+(5/3) m_t^2)\psi(5/3,m_t)}^{1/12}\mu_{30}^{1/4}\,.
\eeq 

A similar estimate for the critical mass accretion rate for settling accretion
can be obtained from a comparison of the characteristic Compton cooling time with the 
convective time at the Alfven radius.

\section{Specific X-ray pulsars}

In this Section, as an illustration of the possible applicability of our model
to real sources, 
we consider three particular slowly rotating moderatly luminous X-ray pulsars: GX 301-2,
Vela X-1, and GX 1+4. The first two pulsars are close to the equilibrium rotation of the neutron star, showing spin-up/spin-down excursions near the equilibrium frequency
(apart from the spin-up/spin-down jumps, which may be, we think, due to episodic 
switch-ons of the strong coupling regime when the toroidal magnetic field component
becomes comparable to the poloidal one, see Section \ref{s:strongcoupling}). The third one, GX 1+4, is a typical example of a pulsar displaying long-term spin-up/spin-down episodes.
During the last 30 years, it shows a steady spin-down with luminosity-(anti)correlated frequency fluctuations (see Gonz\'alez-Gal\'an et al. 2011 for a more detailed discussion). Clearly, this pulsar can not be considered to be in equilibrium.

\subsection{GX 301-2}
GX301--2 (also known as 4U1223--62) is a high-mass X-ray binary, consisting of a neutron star and an early type B optical companion with mass $\simeq 40 M_\odot$ and radius $\simeq 60 R_\odot$. The binary period is 41.5 days (Koh et al. 1997).
The neutron star is a $\sim680$ s X-ray pulsar (White et al. 1976), accreting from the strong wind of its companion ($\dot M_{loss} \sim 10^{-5} \ms$/yr, Kaper et al. 2006). The 
photospheric escape velocity of the wind is $v_{esc}\approx 500$~km/s. The semi-major
axis of the binary system is $a\approx 170 R_\odot$ and the orbital eccentricity $e\approx 0.46$. 
The wind terminal velocity was found by Kaper et al. (2006) to be about 300 km/s, smaller than the photospheric escape velocity.  

GX 301-2 shows strong short-term pulse period variability, which, as in many other wind-accreting pulsars, can be well described by a random walk model (deKool \& Anzer 1993). 
Earlier observations between 1975 and 1984 showed a period of $\sim 700$s while in 1984 the source started to spin up (Nagase 1989). BATSE observed two rapid spin-up episodes, each lasting about 30 days, and it was suggested that the long-term spin-up trend may have been due entirely to similar brief episodes as virtually no net changes in the frequency were found on long time-scales (Koh et al. 1997; Bildsten et al. 1997). The almost 10 years of spin-up were followed by a reversal of spin in 1993 (Pravdo \& Ghosh 2001) after which the source has been continuously spinning down (La Barbera et al. 2005; Kreykenbohm et al. 2004, Doroshenko et al. 2010). Rapid spin-up episodes sometimes appear in the Fermi/GBM data on top of the long-term spin-down trend (Finger et al. 2011). It can not be excluded that these rapid spin-up episodes, as well as the ones observed in the BATSE data, reflect a temporary entrance into the strong coupling regime, as discussed in Section 2.4.1. Cyclotron line measurements (La Barbera et al. 2005) yield the magnetic 
field estimate near the neutron star surface 
$B_0\approx 4.4\times 10^{12}$~G ($\mu=1/2 B_0 R_0^3=2.2\times 10^{30}$~G cm$^3$ for
the assumed neutron star radius $R_0=10$~km).


\begin{figure*}
\includegraphics{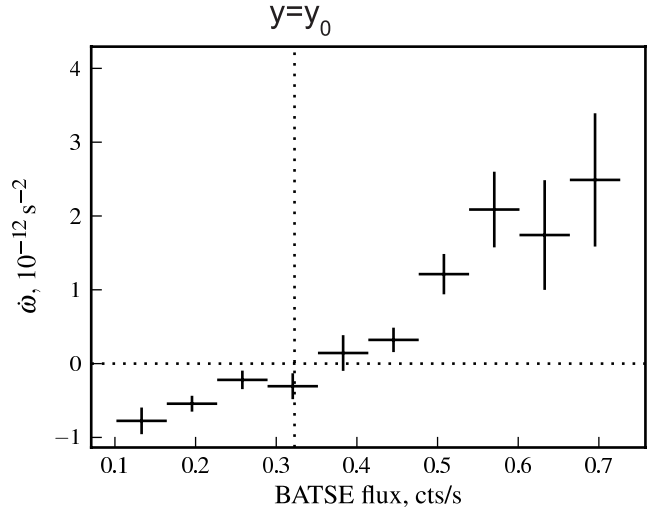}
\caption{Torque-luminosity correlation in GX 301-2, $\dot\omega^*$ as a function of 
BATSE data (20-40 keV pulsed flux) near the equilibrium frequency, see Doroshenko et al. (2010). The assumed X-ray flux at equilibrium (in terms of
the dimensionless parameter $y$) is also shown by the vertical dotted line.}
\label{f:gx301}
 \end{figure*}

In Fig. \ref{f:gx301} we have plotted $\dot\omega^*$ as a function of 
the observed pulsed flux (20-40~keV) according to BATSE data 
(see Doroshenko et al. 2010 for more detail). To obtain the magnetic field
estimate and other parameters (first of all, the coefficient $A$, see 
\Eq{Adet}) as described above in Section \ref{s:magfield}, we need to know 
the value of $\dot M$ and the 
derivative $\partial \dot \omega^*/\partial \dot M$ or $\partial \dot \omega^*/\partial y$.
The estimate of $\dot M$ can be inferred from the X-ray flux provided 
the distance to the source is known, and generally this is a major uncertainty.
We shall assume that near equilibrium a hot quasi-spherical shell exists
in this pulsar, i.e. the accretion rate is $3\times 10^{16}$~g/s, i.e. 
not higher than the critical value $\dot M_*\simeq 4\times 10^{16}$~g/s [\Eq{M*}].
While the absolute value of the mass accretion rate is necessary to 
estimate the magnetic field according to \Eq{mu_eq} (however, the dependence is rather weak, 
$\sim \dot M^{1/3}$), the derivative $\partial \dot \omega^*/\partial y$
can be derived from the $\dot \omega^*$ -- X-ray flux plot, since 
in the first approximation the accretion rate is proportional to the observed pulsed X-ray flux. Near the equilibrium (the torque reversal point with $\dot \omega^*=0$), we find 
from a linear fit in Fig. \ref{f:gx301}   
$\partial \dot \omega^*/\partial y \approx 4 \times 10^{-13}$~rad/s$^2$. 

The obtained parameters ($\mu$, $Z$, $K_1(theta)$, etc.) for this pulsar are listed in Table 1. We note that
the magnetic field estimate resulting from our model for $n=2$ (boldfaced in Table 1) is fairly close to
the value inferred from the cylcotron line measurements. 
We also note that for the case $n=3/2$ the
coupling constants $Z$ and $K_1(\theta)$ turn out to be unrealistically large and
the derived magentic field is very small, suggesting
that assuming anisotropic turbulence is more realistic than
using isotropic turbulence with the viscosity described by the Prandtl law.

\subsection{Vela X-1}

\begin{figure*}
\includegraphics{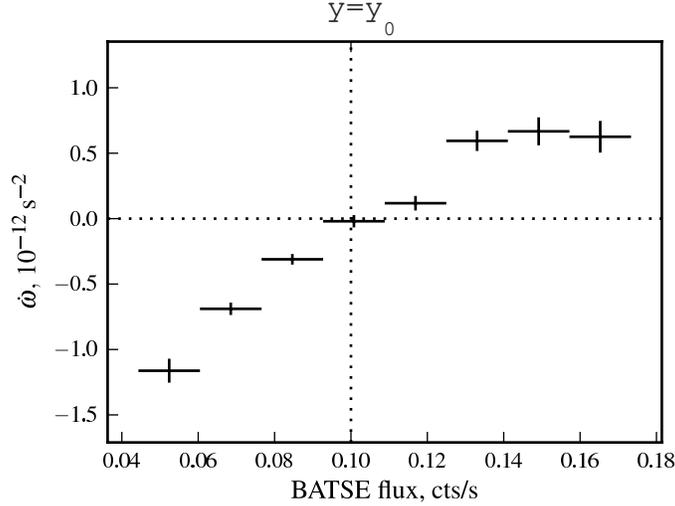}
\caption{The same as in Fig. \ref{f:gx301} for Vela X-1 (Doroshenko 2011,
private communication). }
\label{f:velaX1}
 \end{figure*}
Vela X-1 (=4U 0900-40) is the brightest persistent accretion-powered pulsar in the 20-50 keV energy band with an average luminosity of $L_{x} \approx 4\times10^{36}$erg/s (Nagase 1989).
It consists of a massive neutron star (1.88 $\msun$, Quaintrell et al. 2003) and the B0.5Ib super giant HD 77581, which eclipses the 
neutron star every orbital cycle of $\sim 8.964$ d (van Kerkwijk et al. 1995). The neutron star was discovered as an X-ray pulsar with a spin period of $\sim$283 s (Rappaport 1975), which has remained almost constant since the discovery of the source. The optical companion has a mass and radius of $\sim 23$ $\msun$ and $\sim30$ $R_{sun}$ respectively (van Kerkwijk et al. 1995). The photospheric escape velocity is $v_{esc}\approx 540$~km/s. The orbital separation is $a\approx 50 R_\odot$ and the orbital eccentricity $e\approx 0.1$. 
The primary almost fills its Roche lobe (as also evidenced by 
the presence of elliptical variations in the optical light curve, Bochkarev et al. (1975)). The mass-loss rate from the primary star is $10^{-6}$ $\msun$/yr (Nagase et al. 1986) via a fast wind with a terminal velocity of $\sim $ 1100 km/s (Watanabe et al. 2006), which is typical for this class. 
Despite the fact that the terminal velocity of the wind is rather large, the compactness of the system makes it impossible for the wind to reach this velocity before interacting with the neutron star, so the relative velocity of the wind with respect to neutron star is rather low, $\sim 700$~km/s.

Cyclotron line measurements (Staubert 2003) yields the magnetic field estimate 
$B_0\approx 3\times 10^{12}$~G ($\mu=1.5\times 10^{30}$~G cm$^3$
for the assumed neutron star radius 10~km). We shall assume that in this pulsar  
$\dot M\simeq 3\times 10^{16}$~g/s (again for the existence of the shell to be possible).
In Fig. \ref{f:velaX1} we have plotted $\dot\omega^*$ as a function of 
the observed pulsed flux (20-40~keV) according to BATSE data 
(Doroshenko 2011, private communication). As in the case of GX 301-2, from a
linear fit we
find at the spin-up/spin-down transition point 
$\partial \dot \omega^*/\partial y \approx 5.5\times 10^{-13}$~rad/s$^2$.  

The obtained parameters for Vela X-1 are listed in Table 1.
As in the case of GX 1+4, the magnetic field estimate given by our model 
for an almost iso-angular-momentum rotation law ($n=2$, boldfaced in Table 1) is close
to the value inferred from the cyclotron line measurements.

\subsection{GX 1+4}


GX 1+4 was the first source to be identified as a symbiotic binary containing a neutron star (Davidsen, Malina \& Bowyer 1977). The pulse period is $\sim 140$~s  and the donor is an MIII giant (Davidsen et al. 1977). The system has an orbital period of 1161 days (Hinkle et al. 2006) making it the widest known LMXB by at least one order of magnitude. The donor is far 
from filling its Roche lobe and accretion onto the neutron star is by capture of the stellar wind of the companion. 

The system has a very interesting spin history. During the 1970's it was spinning up at the fastest rate ($\dot \omega_{su} \sim 3.8\cdot 10^{-11}$ rad/s) among the known X-ray pulsars at the time (e.g. Nagase 1989). After several years of non-detections in the early 1980's, it reappeared again, now spinning down at a rate similar in magnitude to that of the previous spin-up.  This spin-reversal has been interpreted in terms of a retrograde accretion disc forming in the system (Makishima et al. 1988, Dotani et al. 1989, Chakrabarty et al. 1997). 
A detailed spin-down history of the source is discussed in the recent paper by Gonz\'alez-Gal\'an et al. (2011).
Using our model this behavior can, however, be readily explained by quasi-spherical accretion.

\begin{figure*}
\includegraphics[width=0.8\textwidth]{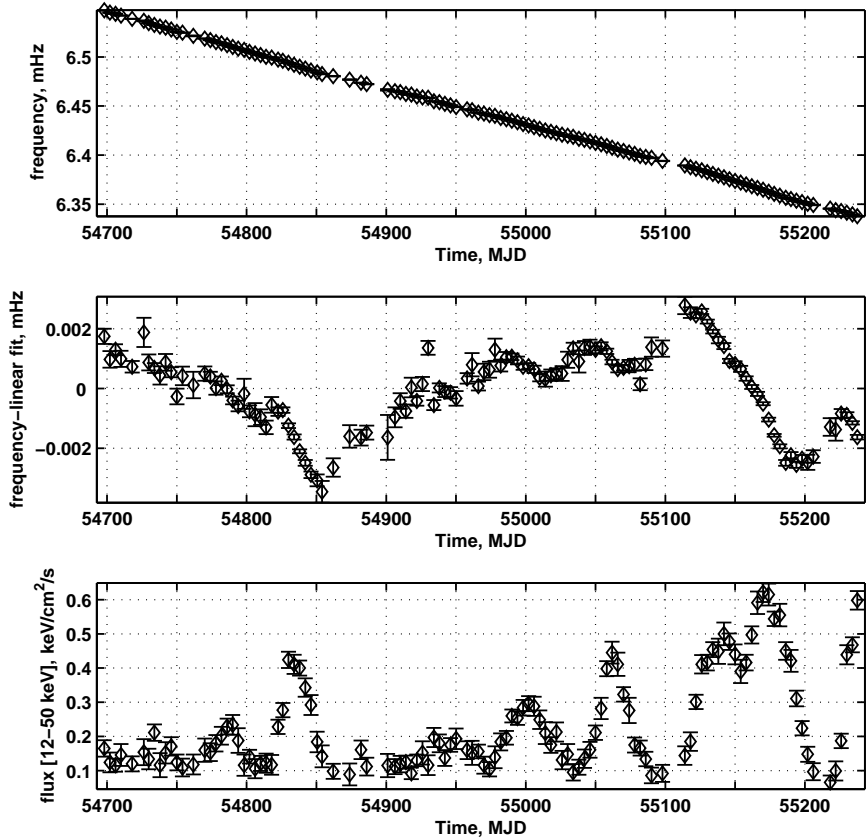}
\caption{Pulsar frequency (upper panel), 
deviation of the frequency from the linear fit (middle panel) 
and  pulsed flux in GX 1+4 from \textit{Fermi GBM} data (M. Finger, private communication;
see also Gonz\'alez-G\'alan et al (2011))}.
\label{f:gx14_f}
 \end{figure*}

As the pulsar in GX 1+4 is not in equilibrium, we cannot directly use our method to 
estimate the magnetic field of the neutron star as described in Section \ref{s:magfield}. However, as GX 1+4 is currently experiencing a long-term spin-down trend, the 
first (spin-up) term in \Eq{sd_eq2} must be smaller than the second (spin-down) term,
which yields a lower limit on the value of the magnetic field (see Table 1). 

Let us use \Eq{variations} to quantitatively 
explain the observed anti-correlation between the pulsar frequency fluctuations and the X-ray flux, as observed in GX 1+4 \citep{g1}. We use a fragment of 
four-day average {\it Fermi}/GBM data on GX 1+4 (M. Finger, private communication;
see also Gonz\'alez-G\'alan et al (2011)). The pulsar is currently observed at the
steady spin-down stage with a mean $\dot \omega_{sd}\approx -2.34\times 10^{-11}$~rad/s
(the upper panel). In the middle panel of Fig. \ref{f:gx14_f}, deviations from
the linear fit are shown. Note that the frequency excursions around the mean value 
is a few microseconds, while the pulsar frequency is much higher, 
a few milliseconds. This means that the frequency derivative is 
negative at all points  within the time interval shown, 
i.e. no occasional spin-ups were observed,
even at the highest X-ray flux levels.

Specifically, let us consider the prominent pulsar frequency change observed between 
MJD 55100 and MJD 55200 for around $\Delta t=80$ days. During this time period, the frequency of the pulsar decreased by $\Delta\omega^*\approx -3.6\times 10^{-5}$~rad (see Fig. \ref{f:gx14_f}). From here we find $\delta \dot \omega^*(obs)=\Delta \omega^*/\Delta t\approx  
-5.2\times 10^{-12}$~rad/s. Thus, the observed fractional change in the pulsar spin-down
rate is $(\delta \dot \omega^*/|\dot\omega^*|)_{obs}\approx -0.2$.

On the other hand, by dividing \Eq{variations} with \Eq{sdy} 
we find the expected relative fluctuations of $\dot \omega^*$
at a given mean accretion rate (or dimensionless X-ray luminosity $y$):
\beq{}
\frac{\delta \dot \omega^*}{|\dot\omega^*|}_{theor}=
\frac{3+2n}{11}\frac{\delta \dot M}{\dot M}
\left[
\frac{1-y^{-\frac{2n-1}{11}}}{1-\myfrac{y_0}{y}^{\frac{2n}{11}}}
\right]
\eeq
From Fig. \ref{f:gx14_f} (bottom panel) 
the range of the fluctuations relative to the mean value 
$(\delta \dot M/\dot M)\simeq (0.6-0.2)/0.4\simeq 1$. Then, 
for the assumed $n=2$, the expected amplitude of the relative 
frequency derivative fluctuation would match the observed value if $y\simeq 0.2-0.3$.
Important is that we find $y<1$, as must be the case for the observed negative 
sign of the torque-luminosity correlations.

Note here 
that the drop in flux-level during about 20 days at around MJD 55140-55160 should be  
translated to a 
decrease in $\dot \omega^*$, which is clearly seen in the upper panel of Fig. \ref{f:gx14_f}.
A more detailed analysis using the entire {\it Fermi} data-set should be performed. 

Further, note that the short-term spin-up episodes, sometimes observed on top
of the steady spin-down behaviour (at about MJD 49700, see Fig. 2 in Chakrabarty et al. (1997)) are 
correlated with an enhancement of the X-ray flux, in contrast to the negative frequency-flux correlation
discussed above. During these short spin-ups, $\dot\omega^*$ is about half the average 
$\dot\omega^*_{su}$ observed during the steady spin-up state of GX 1+4.
The X-ray luminosity during these episodic spin-ups is approximately five times 
larger than the mean X-ray luminosity during the steady spin-down. We remind the reader that 
once $\dot M>\dot M_*$, a free-fall gap appears above the magnetosphere, and the
neutron star can only spin up. When the X-ray flux drops again, the settling accretion 
regime is reestablished and the neutron star resumes its spinning-down.

\begin{table*}
\label{T2}
 \centering
 \caption{Parameters for the pulsars in Section 6. References for the observed spin periods, binary periods and spin down rate (for GX 1+4) are given in the text as well as discussions on the values used for the wind velocities with respect to the neutron star. The parameters Z, K$_1(\Theta)$ and f(u) are defined in Section 2.4.2. Numerical estimates are given for 
dimensionless parameters $\delta=1, \zeta=1$, $\tilde\omega=1$, $\gamma=5/3$ and without turbulence ($m_t=0$). The numbers in boldface are
the preferred values for a near iso-angular-momentum rotation law.} 
 $$
\begin{array}{lcccccc}
\hline
Pulsar & \multicolumn{2}{c}{\rm GX 301-2} & \multicolumn{2}{c}{\rm Vela X-1} 
& \multicolumn{2}{c}{\rm GX 1+4} \\
\hline
\multicolumn{7}{c}{\hbox{Measured parameters}}\\
\hline
P_* {\rm(s)} & \multicolumn{2}{c}{680} & \multicolumn{2}{c}{283} & \multicolumn{2}{c}{140}\\
P_B {\rm(d)} & \multicolumn{2}{c}{41.5} & \multicolumn{2}{c}{8.96} & \multicolumn{2}{c}{1161}\\
v_{w} {\rm(km/s)} & \multicolumn{2}{c}{300} & \multicolumn{2}{c}{700} & \multicolumn{2}{c}{200}\\
\frac{\partial \dot \omega}{\partial y} \arrowvert_{y_{0}} 
& \multicolumn{2}{c}{4\cdot10^{-13}} & \multicolumn{2}{c}{5.5\cdot10^{-13}} & \\
 \dot M_{16}  {\rm(\dot M/10^{16}}) &\multicolumn{2}{c}{3} & \multicolumn{2}{c}{3} & 
\multicolumn{2}{c}{1}\\
\hline
\multicolumn{7}{c}{\hbox{Derived parameters}}\\
\hline
& {\bf n=2} & n=3/2 & {\bf n=2} & n=3/2 & {\bf n=2} & n=3/2\\
\hline
\mu_{30}    & {\bf 2.7} & 0.1 & {\bf 1.8} & 0.16 & {\bf >1.17} &>0.02 \\
f(u) & {\bf 0.42} & 0.57 & {\bf 0.43} & 0.54\\
K_1(\Theta) & {\bf 39} & 3700 & {\bf 36} & 1150\\
Z& {\bf 13} & 910 & {\bf 12} & 300\\
B_t/B_p & {\bf 0.1} & 0.01 & {\bf 0.2} & 0.03\\
R_A{\rm(cm)}& {\bf 2\cdot 10^9} & 3\cdot 10^8 &{\bf 1.6\cdot 10^9}& 4.2\cdot 10^8\\
\omega^*/\omega_K(R_A)& {\bf 0.06} & 0.004 &{\bf 0.1}& 0.01\\
\hline
\end{array}
$$
\end{table*}

\section{Discussion}

\subsection{Physical conditions inside the shell }

For an accretion 
shell to be formed around the neutron star magnetosphere it is necessary that 
the matter crossing the bow shock does not cool down too rapidly and thus starts to fall freely. 
This means that the radiation cooling time $t_{cool}$ must be longer than the characteristic 
time of plasma motion. 

The plasma heats up in the strong shock to the temperature
\beq{T_ps}
T_{ps}=\frac{3}{16}\mu_m\frac{ v_w^2}{\cal R}\approx 1.36\times 10^5 
[\hbox{K}]\myfrac{v_w}{100 \hbox{km/s}}^2\,.
\eeq
The radiative cooling time of the plasma is
\beq{t_cool}
t_{cool}=\frac{3kT}{2\mu_m n_e \Lambda}
\eeq 
where $\rho$ is the plasma density, $n_e=Y_e\rho/m_p$ is the electron number density 
( $\mu_m=0.6$ and $Y_e\approx 0.8$ for fully ionized plasma 
with solar abundance); 
$\Lambda$ is the cooling function which can be approximated as 
\begin{equation}
\label{mcore} \Lambda (T)=\left\{
\begin{array}{l}
0, T<10^4 \, {\rm K} \\
1.0\times 10^{-24} T^{0.55} , 10^4 \, {\rm K} <T<10^5 \, {\rm K}   \\
6.2\times 10^{-19} T^{-0.6} , 10^5 \, {\rm K} <T<4\times 10^7 \, {\rm K}  \\
2.5\times 10^{-27} T^{0.5} , T>4\times 10^7 \, {\rm K}
\end{array}
\right. \label{eqn:lam}
\end{equation}
(Raymond, Cox \& Smith 1976; Cowie,
McKee \& Ostriker 1981).

Compton cooling becomes effective from the radius where the gas temperature $T$,
determined by the hydrostatic formula \Eq{hse_sol}, is lower than the X-ray Compton 
temperature $T_x$. The Compton cooling time (see \Eq{t_comp}) is:
\beq{t_C1}
t_{C}\approx 1060[\hbox{s}] \dot M_{16}^{-1}\myfrac{R}{10^{10}\hbox{cm}}^2\,.
\eeq

Above the
radius where $T_x=T$, Compton heating dominates. 
Taking the actual temperature close to the adiabatic one [\Eq{hse_sol}], 
we find $R_x\approx 2\times 10^{10}$~cm. 
We note that both the Compton 
and photoionization heating processes are controlled by the photoionization parameter
$\xi$ (Tarter et al. 1969, Hatchett et al. 1976)
\beq{ksi}
\xi=\frac{L_x}{n_eR^2}\,.
\eeq 
In most part of the accretion flux, $n\sim R^{-3/2}$, so $\xi\sim R^{-1/2}$ and
independent of the X-ray luminosity through the mass continuity equation. 
For characteristic values we find:
\beq{kxi_n}
\xi\approx 5\times 10^5 f(u) R_{10}^{-1/2}\,.
\eeq 
If Compton processes were effective everywhere, this high value of the parameter $\xi$
would imply that the plasma is Compton-heated up to keV-temperatures out to very large 
distances $\sim 10^{12}$~cm. However, at large distances the Compton heating time 
becomes longer than the characteristic time of gas accretion:
\beq{}
\frac{t_{C}}{t_{accr}}=\frac{t_{C}f(u)u_{ff}}{R}\approx 20 f(u) \dot M_{16}^{-1}R_{10}^{1/2}\,,
\eeq 
which shows that Compton heating is ineffective. The gas temperature is determined
by photoionization heating only and 
the gas can only be heated up to $T_{max}\approx 5\times 10^5$~K (Tarter et al. 1969),
which is substantially lower than $T_x\sim 3$~keV. The sound velocity corresponding to $T_{max}$ is approximately 80 km/s. 

The effective gravitational capture radius corresponding to the sound velocity of the gas in the photoionization-heated zone is 
\beq{R_BC}  
R_{B}^*=\frac{2GM}{c_s^2}=\frac{2GM}{\gamma{\cal R} T_{max}/\mu_m}\approx 3.5\times 10^{12}\hbox{cm}
\myfrac{T_{max}}{5\times 10^5\hbox{K}}^{-1}\,.
\eeq
Everywhere up to the bow shock photoionization keeps the temperature at a value
$\simeq T_{max}$. 
 
If the stellar wind velocity exceeds $80$~km/s, 
a standard bow shock is formed at the Bondi radius with a post-shock temperature given by \Eq{T_ps}. If the stellar wind velocity is lower than this value, the shock disappears and quasi-spherical accretion occurs from $R_B^*$. 

The photoionization heating time at the effective Bondi radius $3\times 10^{12}$~cm  is 
\beq{}
t_{pi}\approx \frac{(3/2)kT_{max}/\mu_m}{(h\nu_{eff}-\zeta_{eff})n_\gamma \sigma_{eff}c}
\approx 2\times 10^4 [\hbox{s}] \dot M_{16}^{-1}\,.
\eeq
(here $h\nu_{eff}\sim 10$~keV is the characteristic photon energy, $\zeta$ is the effective photoionization potential, $\sigma_{eff}\sim 10^{-24}$~cm$^2$ is the typical photoionization cross-section, $n_\gamma=L/(4\pi R^2 h\nu_{eff} c)$ is the photon number
density). The photoionization to accretion time ratio at the effective Bondi radius is then
\beq{}
\frac{t_{pi}}{t_{accr}}\approx 0.07 f(u) \dot M_{16}^{-1}\,.
\eeq

At wind velocities $v_w>80$ km/s the bow shock stands at the classical Bondi radius $R_B$
inside the effective Bondi radius
$R_B^*$
determined by \Eq{R_BC}. The cooling time of the shocked
plasma at $R_B$ expressed through the wind velocity $v_w$ is: 
\beq{t_cool1}
t_{cool}\approx 4.7\times 10^4 [\hbox{s}]\dot M_{16}^{-1} v_7^{0.2}\,.
\eeq 
The photoionization heating time in the post-shock region can also be expressed through the
stellar wind velocity: 
\beq{}
t_{pi}\approx 3.5 \times 10^4 [\hbox{s}]\dot M_{16}^{-1} v_7^{-4}\,.
\eeq
The comparison of these two timescales implies that at low velocities radiative cooling is
important and the regime of free-fall accretion with conservation of specific angular momentum is realized. 

So, at low wind velocities the plasma cools down and starts to fall freely. As the cold plasma approaches
the gravitating center, photoionization heating becomes important and rapidly heats up the plasma to $T_{max}\approx 5\times 10^5$~K. Should this occur at a radius where $T_{max}<GM/({\cal R}R)$, the plasma continues its free fall down to the magnetosphere, still with the temperature $T_{max}$, with the
subsequent formation of a shock above the magnetosphere. However,
if $T_{max}$ is above the adiabatic temperatures at this radius, the settling accretion regime 
will be established even for low wind velocities. 

For high-wind stellar velocities $v_w\gtrsim 100$~km/s, the post-shock temperature is higher than $T_{max}$, photoionization is unimportant, and the settling accretion regime 
is established if the radiation cooling time is longer than the accretion time. 
From a comparison of these timescales, we find the critical accretion rate as a function 
of of the wind velocity below which the settling accretion regime is possible:
\beq{}
\dot M_{16}^{**}\lesssim 0.12 v_7^{3.2}\,.
\eeq

Here we stress the difference of the critical acccretion rate $\dot M^{**}$ from 
$\dot M^*$ derived earlier. At $\dot M>\dot M^{**}$, the plasma rapidly cools down 
in the gravitational capture region and free-fall accretion begins (unless photoionization
heats up the plasma above the adiabatic value at some radius), while at $\dot M>\dot M^*
\simeq 4\times 10^{16}$~g/s 
determined by \Eq{M*} a free-fall gap appears immediately above the 
neutron star magnetosphere. 

\subsection{On the possibility of the propeller regime}

 The very slow rotation of the neutron stars in GX 1+4, GX 301-2 and Vela X-1 
($\omega^* (R_A)<\omega_K(R_A)$) 
makes it hard to establish the propeller regime 
where matter is ejected with parabolic velocities from the magnetosphere during spin-down episodes.

Let us therefore start with estimating the important 
ratio of viscous tensions ($\sim B_tB_p$) 
to the gas pressure ($\sim B_p^2$) at the magnetospheric boundary. This ratio is proportional to 
$B_t/B_p$ (see \Eq{BtBpnum}) and is always much smaller than 1 (see Table 1),
i.e. only large-scale convective  
motions with the characteristic hierarchy of 
eddies scaled with radius can be established in the shell. 
When $\omega^*>\omega_K(R_A)$, the propeller regime (without accretion) 
must set in. In that case the maximum possible braking torque is $\sim -\mu^2/R_A^3$
due to the strong coupling between the plasma and the magnetic field.
Note that in the propeller state, interaction of the plasma with the magnetic field
is in the strong coupling regime, i.e. where the toroidal 
magnetic field component $B_t$ is comparable to the poloidal one $B_p$. 
It can not be excluded that a hot iso-angular-momentum envelope could exist 
in this case as well, which would then remove angular momentum from the rotating magnetosphere. If the characteristic cooling time of the gas in the envelope is short
in comparison to the falling time of matter, the shell disappears and 
one can expect the formation of a `storaging' thin Keplerian disc around the 
neutron star magnetosphere (Sunyaev \& Shakura 1977). There is no accretion of matter through such a disc. It 
only serves to remove angular momentum from the magnetosphere. 

\subsection{Effects of the hot shell on the X-ray energy and power spectrum}

The spectra of X-ray pulsars are dominated by emission generated in 
the accretion column. The hot optically thin shell produces its own thermal emission, but even if 
all gravitational energy were released in the shell, the ratio of the X-ray luminosity 
from the shell to that of the accretion column would be about the ratio of the
magnetosphere radius to the NS radius, i.e. one percent or less. In reality, it is much 
smaller. The shell should scatter X-ray radiation from the accretion column, but 
for this effect to be substantial, the Comptonization parameter $y$ must be of 
the order of one. The Thomson depth in the shell is, however, very small. Indeed, from the mass continuity equation 
and \Eq{RA} for the Alfven radius and \Eq{fu} for the factor $f(u)$, we get:
$$
\tau_T=\int_{R_A}^{R_B}n_e(R)\sigma_T dR
\approx 3.2\times 10^{-3} \dot M_{16}^{8/11}\mu_{30}^{-2/11}\,.
$$
Therefore, for the temperature near the magnetosphere [\Eq{hse_sol}] the parameter $y$ is 
$$
y=\frac{4kT}{m_ec^2}\tau_T\approx 2.4\times 10^{-3}\,.
$$
This means that the X-ray spectrum of the accretion column should not be
significantly affected by scattering in the hot shell.

The large-scale convective motions in the shell introduce 
an intrinsic time-scale of the order of
the free-fall time that could give rise to 
features (e.g. QPOs) in the power spectrum of variability. 
QPOs were reported in some X-ray pulsars 
(see Marykutty et al. 2010 and references therein).
However, the expected frequency of the QPOs arising
in our model would be of the order of mHz, much shorter than those reported.

A stronger effect can be  
the appearance of a dynamical instability of the shell on 
this time scale due to increased Compton cooling
and hence increased mass accretion rate in the shell. This  
may result in a complete collapse of the shell
resulting in an X-ray outburst with duration similar to 
the free-fall time scale of the shell
($\sim 1000$~s). Such a transient behaviour 
is observed in supergiant fast X-ray transients (SFXTs) (see Ducci et al. 2010). 
This interesting issue depends on the specific parameters
of the shell and needs to be further investigated.

\subsection{Can accretion discs (prograde or retrograde) 
be present in these pulsars?}

The analysis of real pulsars carried out earlier suggested that in a convective shell an 
iso-angular-momentum distribution is the most plausible. Therefore,   
we shall below consider only this case, i.e. the rotation law $\omega\sim R^{-2}$. 
As follows from \Eq{sd_eq1}, at $\dot \omega^*=0$ the equilibrium angular 
frequency of the neutron star is
\beq{equilib}
\omega^*_{eq}=\omega_B\frac{1}{1-z/Z}\myfrac{R_B}{R_A}^2\,.
\eeq
We stress that such an equilibrium in our model is possible only 
when a shell is present. At high accretion rates 
$\dot M>\dot M_*\simeq 4\times 10^{16}$~g/s accretion proceeds 
in the free-fall regime (with no shell present).

Using \Eq{mu_eq}, 
the equlibrium period for quasi-spherical settling accretion can be recasted to the form
\beq{P_eq}
P_{eq}\approx 1000 [\hbox{s}]\mu_{30}^{12/11}\myfrac{P_b}{10\hbox{d}}
\myfrac{\zeta}{(1+(5/3) m_t^2)\psi(5/3,m_t)\dot M_{16}}^{4/11}\myfrac{v_8}{\sqrt{\delta}}^4
\frac{(1-z/Z)}{\tilde \omega(1+(5/3)m_t^2)}\,.
\eeq

For standard disc accretion, the equilibrium period is
\beq{P_eqd}
P_{eq,d}\approx 7\hbox{s} \mu_{30}^{6/7}\dot M_{16}^{-3/7}\,,
\eeq
and the long periods observed in some X-ray pulsars can be explained 
assuming a very high magnetic field of the neutron star.  
Retrograde accretion discs are also discussed in the literature (see, e.g., 
Nelson et al. (1997) and references therein). Torque reversals produced by 
prograde/retrograde discs can in principle lead to very long periods for X-ray pulsars
even with standard magnetic fields. Retrograde discs can be formed due to inhomogeneities in 
the captured stellar wind (Ruffert 1997, 1999). This might be the case
at high accretion rates when hot quasi-spherical shell cannot exist. 
In the case of GX 1+4, however, it is highly unlikely to observe a retrograde disk 
on a time scale much longer than the orbital period (see a more detailed discussion 
of this issue in 
Goz\'alez-Gal\'an et al. (2011)). In the case of GX 301-2 and Vela X-1, the observed positive
torque-luminosity correlation (see Figs. \ref{f:gx301} and \ref{f:velaX1}) rules
out a retrograde disc as well.

To conclude the discussion, we should mention that 
real systems (including those considered here) demonstrate 
a complex quasi-stationary behaviour with dips, outbursts, etc. 
These considerations are beyond the scope of this paper and definitely 
deserve further observational and theoretical studies.

\section{Conclusions}

In this paper we have presented a theoretical model for quasi-spherical 
subsonic accretion onto slowly rotating magnetized 
neutron stars. 
In this model the accreting matter is gravitationally 
captured from the stellar wind of the optical companion and subsonically 
settles down onto the rotating magnetosphere forming an extended quasi-static shell.
This shell mediates the angular momentum removal from the rotating neutron star 
magnetosphere during spin-down states by large-scale convective motions. 
A detailed analysis and comparison with observations of two specific X-ray pulsars GX 301-2 and Vela X-1 demonstrating torque-luminosity correlations near the
equilibrium neutron star spin period shows that most likely 
strongly anisotropic convective motions are established, with an 
almost iso-angular-momentum distribution of rotational velocities $\omega\sim R^{-2}$.
A statistical analysis of long-period X-ray pulsars with Be-components in SMC
(Chashkina \& Popov 2011)
also favored the rotation law $\omega\sim R^{-2}$.
The accretion rate through the shell is determined by the ability of the plasma to enter the magnetosphere. The settling regime of accretion which allows angular momentum removal
from the neutron star magnetosphere can be realized for moderate accretion rates 
$\dot M< \dot M_*\simeq 4\times 10^{16}$~g/s. At higher accretion rates a free-fall gap 
above the neutron star magnetosphere appears due to rapid Compton cooling, and accretion 
becomes highly non-stationary. 

From observations of the spin-up/spin-down rates (the angular rotation frequency derivative 
$\dot \omega^*$, or  $\partial\dot\omega^*/\partial\dot M$ near the torque reversal) 
of long-period X-ray pulsars with known orbital periods it is possible 
to determine the main dimensionless parameters of the model, as well as to estimate the magnetic field of the neutron star. Such an analysis revealed good agreement between the 
magnetic field estimates in the pulsars 
GX 301-2 and Vela X-1 obtained using our model and derived from the cyclotron 
line measurements.    

In our model, long-term spin-up/spin-down as observed in some X-ray pulsars can be quantitatively explained by a change in the mean mass accretion rate onto the neutron star (and the corresponding mean X-ray luminosity). Clearly, these
changes are related to the stellar wind properties. 

The model also predicts the specific behaviour 
of the variations in $\delta \dot \omega^*$, observed on top of a steady spin-up or spin-down,
as a function of mass accretion rate fluctuations $\delta \dot M$. There is a critical accretion rate $\dot M_{cr}$ below which an anti-correlation of $\delta \dot \omega^*$ with $\delta \dot M$ should occur (the case of 
GX 1+4 at the steady spin-down state currently observed), and above which 
$\delta \dot\omega^*$ should correlate with $\delta \dot M$ fluctuations (the case 
of Vela X-1, GX 301-2, and GX 1+4 in the steady spin-up state). The model explains quantitatively the relative amplitude and the sign 
of the observed frequency fluctuations in GX 1+4.  

\appendix
\section{The structure of a quasi-spehrical accreting rotating shell}

\subsection{Basic equations}
Let us first write down the Navier-Stokes equations in spherical coordinates $R, \theta, \phi$. Due to the huge Reynolds numbers in the shell ($\sim 10^{15}-10^{16}$ for a 
typical accretion rate of $10^{17}$~g/s and magnetospheric radius $\sim 10^8$~cm), there must be turbulence.  
In this case the Navier-Stokes equations are usually called the Reynolds equations.
In the general case, the turbulent viscosity can depend on the coordinates,
so the equations take the form:

1. Mass continuity equation:
\beq{mass_cont}
   \frac{\partial \rho}{\partial t}+ 
\frac{1}{R^2}\frac{\partial}{\partial R}\left(R^2 \rho u_R\right)+ \frac{1}{R \sin\theta}\frac{\partial}{\partial \theta}\left(\sin\theta\, \rho u_\theta\right) + \frac{1}{R \sin\theta}\frac{\partial \rho u_\phi}{\partial \phi}  = 0. 
\eeq

2. The $R$-component of the momentum equation: 
\beq{v_R}
\displaystyle\frac{\partial u_R}{\partial t} + u_R \displaystyle\frac{\partial u_R}{\partial R} + \displaystyle\frac{u_{\theta}}{R} \displaystyle\frac{\partial u_R}{\partial \theta}+ \displaystyle\frac{u_{\phi}}{R \sin\theta} \displaystyle\frac{\partial u_R}{\partial \phi}  - \displaystyle\frac{u_{\phi}^2 + u_{\theta}^2}{R} = 
-\displaystyle\frac{GM}{R^2}+N_R
\eeq

3. The $\theta$-component of the momentum equation: 
\beq{v_theta}    
\displaystyle\frac{\partial u_{\theta}}{\partial t} 
+ u_R \displaystyle\frac{\partial u_{\theta}}{\partial R}
+ \displaystyle\frac{u_{\theta}}{R} \displaystyle\frac{\partial u_{\theta}}{\partial \theta}  
+ \displaystyle\frac{u_{\phi}}{R \sin\theta} \displaystyle\frac{\partial u_{\theta}}{\partial \phi} 
+ \displaystyle\frac{u_R u_{\theta} - u_{\phi}^2 \cot\theta}{R} = 
N_\theta
\eeq

4. The $\phi$-component of the momentum equation: 
\beq{v_phi}
\displaystyle\frac{\partial u_{\phi}}{\partial t} 
+ u_R \displaystyle\frac{\partial u_{\phi}}{\partial R} 
+ \displaystyle\frac{u_{\theta}}{R} \displaystyle\frac{\partial u_{\phi}}{\partial \theta} 
+ \displaystyle\frac{u_{\phi}}{R \sin\theta} \displaystyle\frac{\partial u_{\phi}}{\partial \phi} 
+ \displaystyle\frac{u_R u_{\phi} + u_{\phi} u_{\theta} \cot\theta}{R} =
N_\phi  
\eeq

Here the force components (including viscous force and gas pressure gradients) read:
\beq{N_R}
\rho N_R=\frac{1}{R^2}\frac{\partial}{\partial R}\left( R^2 W_{RR}\right)
+\frac{1}{\sin\theta\,R}\frac{\partial}{\partial \theta}\left(  W_{R\theta}\sin\theta\right)
+\frac{1}{\sin\theta\,R}\frac{\partial}{\partial \phi}W_{R\phi}-\frac{W_{\theta\theta}}{R}
-\frac{W_{\phi\phi}}{R}
\eeq

\beq{N_theta}
\rho N_\theta=\frac{1}{R^2}\frac{\partial}{\partial R}\left( R^2 W_{\theta R}\right)
+\frac{1}{\sin\theta\,R}\frac{\partial}{\partial \theta}\left(  W_{\theta\theta}\sin\theta\right)
+\frac{1}{\sin\theta\,R}\frac{\partial}{\partial \phi}W_{\theta\phi}-\cot\theta\frac{W_{\theta\theta}}{R}
\eeq

\beq{N_phi}
\rho N_\phi=\frac{1}{R^3}\frac{\partial}{\partial R}\left( R^3 W_{\phi R}\right)
+\frac{1}{\sin\theta\,R}\frac{\partial}{\partial \theta}\left(  W_{\phi\theta}\sin\theta\right)
+\frac{1}{\sin\theta\,R}\frac{\partial}{\partial \phi}W_{\phi\phi}
\eeq

The components of the stress tensor include a contribution from both the gas pressure $P_g$ 
(assumed to be isotropic) and the turbulent pressure $P^t$ (generally anisotropic). In their definition we shall follow 
the classical treatment by Landau and Lifshitz (1986), but with the inclusion of 
anisotropic turbulent pressure:
\beq{}
W_{RR}=-P_g-P_{RR}^t+2\rho\nu_t\frac{\partial u_R}{\partial R}-\frac{2}{3}\rho\nu_t{\rm div} {\bf u}
\eeq

\beq{}
W_{\theta\theta}=-P_g-P_{\theta\theta}^t+2\rho\nu_t\left(\frac{1}{R}\frac{\partial u_\theta}{\partial \theta}
+\frac{u_R}{R}\right)-\frac{2}{3}\rho\nu_t{\rm div} {\bf u}
\eeq

\beq{}
W_{\phi\phi}=-P_g-P_{\phi\phi}^t+
2\rho\nu_t\left(\frac{1}{R\sin\theta}\frac{\partial u_\phi}{\partial \phi}
+\frac{u_R}{R}+\frac{u_\theta\cot\theta}{R}\right)-\frac{2}{3}\rho\nu_t{\rm div} {\bf u}
\eeq

\beq{}
W_{R\theta}=\rho\nu_t\left(\frac{1}{R}\frac{\partial u_R}{\partial \theta}
+\frac{\partial u_\theta}{\partial R}-\frac{u_\theta}{R}\right)
\eeq

\beq{}
W_{\theta\phi}=\rho\nu_t\left(\frac{1}{R\sin\theta}\frac{\partial u_\theta}{\partial \phi}
+\frac{1}{R}\frac{\partial u_\phi}{\partial\theta}-\frac{u_\phi\cot\theta}{R}\right)
\eeq

\beq{}
W_{R\phi}=\rho\nu_t\left(\frac{1}{R\sin\theta}\frac{\partial u_R}{\partial \phi}
+\frac{\partial u_\phi}{\partial R}-\frac{u_\phi}{R}\right)
\eeq
In our problem the anistropy of the turbulence is such that $P_{RR}^t=P_\parallel^t$, 
$P_{\theta\theta}^t=P_{\phi\phi}^t=P_\perp^t$. The turbulent pressure components can be 
expressed through turbulent Mach numbers and will be given in Appendix D.

${\rm div}{\bf u}$ in spherical coordinates is:
\beq{}
{\rm div}{\bf u}=
\frac{1}{R^2}\frac{\partial}{\partial R}\left(R^2 u_R\right)
+ \frac{1}{R \sin\theta}\frac{\partial}{\partial \theta}\left(\sin\theta\, u_\theta\right) + \frac{1}{R \sin\theta}\frac{\partial u_\phi}{\partial \phi}. 
\eeq

\subsection{Symmetries of the problem}

We shall consider axially-symmetric ($\displaystyle\frac{\partial}{\partial \phi}=0$), 
stationary ($\displaystyle\frac{\partial}{\partial t}=0$), and only radial accretion
 ($u_\theta=0$). Under these conditions, from the continuity equation \Eq{mass_cont} we obtain:
\beq{dotM}
\dot M= 4\pi R^2\rho u_R=const\,.
\eeq
The constant here is determined from the condition of plasma leakage through the magnetosphere.

Let us rewrite the Reynolds equations under the above assumptions.
The $R$-component of the momentum \Eq{v_R} equation becomes:
\beq{v_R1}
 \rho\left(u_R \displaystyle\frac{\partial u_R}{\partial R} 
  - \displaystyle\frac{u_{\phi}^2}{R}\right) 
= 
-\rho \displaystyle\frac{GM}{R^2}+
\frac{1}{R^2}\frac{\partial}{\partial R}\left( R^2 W_{RR}\right)
+\frac{1}{\sin\theta\,R}\frac{\partial}{\partial \theta}\left(  W_{R\theta}\sin\theta\right)
-\frac{W_{\theta\theta}}{R}
-\frac{W_{\phi\phi}}{R}
\eeq
The $\theta$-component of the momentum equation: 
\beq{v_theta1}    
-\rho\frac{ u_{\phi}^2 \cot\theta}{R} 
=
\frac{1}{R^2}\frac{\partial}{\partial R}\left( R^2 W_{\theta R}\right)
+\frac{1}{\sin\theta\,R}\frac{\partial}{\partial \theta}\left(  W_{\theta\theta}\sin\theta\right)
-\cot\theta\frac{W_{\theta\theta}}{R}
\eeq
The $\phi$-component of the momentum equation: 
\beq{v_phi1}
\rho\left(u_R \displaystyle\frac{\partial u_{\phi}}{\partial R} 
+ \displaystyle\frac{u_R u_{\phi}}{R}\right) 
=\frac{1}{R^3}\frac{\partial}{\partial R}\left( R^3 W_{\phi R}\right)
+\frac{1}{\sin\theta\,R}\frac{\partial}{\partial \theta}\left(  W_{\phi\theta}\sin\theta\right)
\eeq

The components of the stress tensor with anisotropic turbulence 
take the form:

\beq{W_RR}
W_{RR}=-P_g-P_{\parallel}^t-\frac{4}{3}\rho\nu_t\left(\frac{u_R}{R}-\frac{\partial u_R}{\partial R}\right)
\eeq

\beq{W_tt}
W_{\theta\theta}=-P_g-P_{\perp}^t+\frac{2}{3}\rho\nu_t\left(\frac{u_R}{R}-\frac{\partial u_R}{\partial R}\right)
\eeq

\beq{W_pp}
W_{\phi\phi}=-P_g-P_{\perp}^t+\frac{2}{3}\rho\nu_t\left(\frac{u_R}{R}-\frac{\partial u_R}{\partial R}\right)
\eeq

\beq{W_rt}
W_{R\theta}=\rho\nu_t\frac{1}{R}\frac{\partial u_R}{\partial \theta}
\eeq

\beq{W_tf}
W_{\theta\phi}=\rho\nu_t\left(
\frac{1}{R}\frac{\partial u_\phi}{\partial\theta}-\frac{u_\phi\cot\theta}{R}\right)
\eeq

\beq{W_rf}
W_{R\phi}=\rho\nu_t\left(\frac{\partial u_\phi}{\partial R}-\frac{u_\phi}{R}\right)
\eeq

Generally, there are two separate cases -- with isotropic turbulence and anisotropic turbulence, which shall be treated differently. First we consider the simplest case of isotropic
turbulence with Prandtl law for turbulent viscosity.
The more general case of anisotropic turbulence will be discussed separately in Appendix B.

\subsection{Viscosity prescription for isotropic turbulence}

We consider an axisymmetric flow with a very large Reynolds number. By generalizing the 
Prandtl law for the turbulent velocity obtained for plane parallel flows, 
the turbulent velocity scales as 
$u_t\sim l_t R(\partial\omega /\partial R)$. From the similarity laws 
of gas-dynamics we assume $l_t\sim R$,  so
\beq{u_t}
u_t=C_1R^2\left|\frac{\partial\omega }{\partial R}\right|\,.
\eeq
We note that in our case the turbulent velocity is determined by convection, so 
$u_t\lesssim 0.5 u_{ff}$ (see Appendix D). 
This implies that the constant $C_1$ scales as 
\beq{}
C_1\sim u_t/\langle u_\phi \rangle, 
\eeq
and can be very large since $\langle u_\phi \rangle \ll u_t$. 
So the turbulent viscosity coefficient reads:
\beq{nut}
\nu_t=\langle u_t l_t\rangle=C_2C_1 R^3\left|\frac{\partial \omega}{\partial R}\right|
\eeq
Here $C_2\approx 1/3$ is a numerical factor originating from statistical averaging.
Below we shall combine $C_1$ and $C_2$ into the new coefficient $C=C_1C_2$, 
which can be much larger than one.  Due to interaction of 
convection cells in the sheared flux the amplitude of turbulent motions
is thus much larger than the average rotation velocity.

\subsection{The angular momentum transport equation and the rotation law inside the shell}

 A similar problem (that of a rotating sphere in a viscous fluid) was solved in Landau and Lifshitz (1986). It was found that the variables are separated and 
$u_{\phi}(R,\theta)= U_\phi(R)\sin\theta$. Note that the angular velocity 
$\omega(R)=U_\phi(R)/R$ is independent of $\theta$.  
Our problem is different from that of the sphere in a viscous fluid in several respects: 1) there is a force of gravity present, 2) the turbulent viscosity varies with $R$ and can 
in principle depend on $\theta$, and 3) there is radial motion of matter (accretion). These differences lead, 
as will be shown below, to the radial dependence 
$U_\phi (R)\propto R^{-1/2}$. 
(We recall that for a rotating sphere in a viscous fluid $U_\phi\propto R^{-2}$). 

Let us start with solving \Eq{v_phi1}. First, we note that for 
$u_\phi(\theta)\sim \sin\theta$, according to \Eq{W_tf}, $W_{\theta\phi}=0$. 
Next, making use of the continuity equation \Eq{dotM} and the definition of angular velocity, 
we rewrite \Eq{v_phi1} in the form of angular momentum transfer by viscous forces:
\beq{amt}
\frac{\dot M}{R}\frac{\partial}{\partial R}\omega R^2 = \frac{4\pi}{R}\frac{\partial}{\partial R}R^3W_{R\phi}
\eeq

Now we can integrate \Eq{amt} with respect to $R$ to get  
\beq{amt00}
\dot M \omega R^2= 4\pi  R^3 W_{R\phi} +D\,,
\eeq
where $D$ is an integration constant.
In the case of isotropic turbulence, 
the viscous stress component (force per unit area) from \Eq{W_rf} is  
\beq{W_rf1}
W_{R\phi}=\rho \nu_t R \frac{\partial \omega}{\partial R}
\eeq
so we obtain
\beq{amt1}
\dot M \omega R^2= 4\pi \rho \nu_t R^4 \frac{\partial \omega}{\partial R}+D\,.
\eeq

This equation for angular mometum transport by turbulent viscosity is similar to that 
for disc accretion (Shakura and Sunyaev 1973), but different due to spherical symmetry of the problem under consideration. 

The left part of  \Eq{amt1} is simply advection of specific 
angular momentum averaged over the sphere ($1/2\int_0^\pi\omega R^2 \sin^2\theta \sin\theta d\theta=1/3 \omega R^2$) 
by the average motion toward the gravitational center (accretion). $\dot M$ is negative 
as well as $\frac{\partial \omega}{\partial R}$. 
The first term on the right 
describes transport of angular momentum outwards by turbulent viscous forces. 

The constant $D$ is determined from the equation
\beq{D1}
D=K_1(\theta)K_2 \frac{\mu^2}{R_A^3}\frac{\omega_m-\omega^*}{\omega_K(R_A)}
\eeq
(see \Eq{sd1} in the text). We consider accretion onto a magnetized neutron star. 
 When $D<0$, the advection term in the left part of \Eq{amt1} 
dominates over viscous angular momentum transfer outwards. Oppositely,
when $D>0$, the viscous term in the right part of \Eq{amt1} dominates. 
In the case of $\dot M=0$ (no plasma enters the magnetosphere), 
there is only angular momentum transport outwards by viscous forces.

Now let us rewrite \Eq{D1} in the form
\beq{D2}
D=K_1(\theta)K_2
\frac{\mu^2}{R_A^6}R_A^3\frac{\omega_m-\omega^*}{\omega_K(R_A)}
\eeq
and use the pressure balance condition
\beq{}
P(R_A)=P_g(R_A)(1+\gamma m_t^2)=
\frac{B^2(R_A)}{8\pi}=
\frac{K_2}{2\pi}\frac{\mu^2}{R_A^6}\,.
\eeq
Using the mass conitnuity equation in the form 
\[
|\dot M|=4\pi R^2\rho f(u)\sqrt{GM/R}\,,
\]
and the expression for the gas pressure \Eq{P(RA)}, 
we write the integration constant $D/|\dot M|$ in the form
\beq{}
\frac{D}{|\dot M|}=K_1\frac{(\gamma-1)}{\gamma}\psi(\gamma, m_t)
\frac{(\omega_m-\omega^*)R_A^2}{2\sqrt{2}f(u)}(1+\gamma m_t^2)\,.
\eeq

Let us consider the case 
near the neutron star rotation equilibrium $\dot \omega^*=0$. In this case according to 
\Eq{sd_eq} 
\beq{do}
\omega_m-\omega^*=-\frac{z}{Z}\omega^*\,,
\eeq
so using definition $Z$ [\Eq{Zdef}], we obtain:
\beq{Deq}
\frac{D}{|\dot M|}=-zR_A^2 \omega^*\,.
\eeq
We emphasize that the value of the constant $D$ is fully determined by 
the dimensionless specific angular momentum of matter at the Alfven radius $z$.

\subsection{Angular rotation law}


Let us use \Eq{amt1} to find the rotation law $\omega(R)$. 
At large distances $R\gg R_A$ (we remind the reader that $R_A$ is the bottom radius of the shell)
the constant $D$ is small relative to the other terms, so we can set $D\approx 0$.
Thus, to obtain the rotation law we shall neglect this constant in the right part
of \Eq{amt1}. Next, we substitute \Eq{nut} 
and make use of
the solution for the density (which, as we shall show below, remains the same as in the hydrostatic solution)  
\beq{rho_R}
\rho(R)=\rho(R_A)\myfrac{R_A}{R}^{3/2}\,
\eeq
in \Eq{amt1} to obtain
\beq{amt2}
\left|\dot M\right| \omega R^2= 4\pi \rho(R_A) 
\myfrac{R_A}{R}^{3/2}CR^7 \myfrac{\partial \omega}{\partial R}^2\,.
\eeq
After integrating this equation, we find
\beq{sqrtomega}
2\omega^{1/2}=\pm \frac{4}{3}\frac{K^{1/2}}{R^{3/4}}+D_1
\eeq
where 
\beq{}
K=\frac{|\dot M|}{4\pi \rho(R_A) C R_A^{3/2}}
\eeq
and $D_1$ is some integration constant. In \Eq{sqrtomega} we use only the positive 
solution (the minus sign with constant $D_1>0$ would correspond to 
a solution with the angular velocity growing outwards, which is possible if
the pulsar rotation is zero). 
If $D_1\ne 0$, at large $R\gg R_A$ (in the zone close to the bow shock) 
the solid
body rotation law would lead to $\omega\to const\approx \omega_B$. (However, we remind the reader
that our discussion is not applicable close to the bow shock region.) 
At small distances from the Alfvenic surface the effect of
this constant is small and we shall neglect it in the calculations below. Then we find 
\beq{omega}
\omega(R)=\frac{4}{9}\frac{|\dot M|}{4\pi \rho(R_A)CR_A^3}\myfrac{R_A}{R}^{3/2}
\eeq
i.e. the quasi-Keplerian law $\omega(R)=\omega_m(R_A/R)^{3/2}$. 
The constant $\omega_m$ in the solution given by \Eq{omega}
is obtained after substituting $\dot M$ from the continuity equation 
at $R=R_A$ into 
\Eq{omega}:
\beq{omega_m}
\omega_m\equiv \tilde \omega \omega(R_A)=\frac{4}{9}\tilde \omega \frac{|u_R(R_A)|}{CR_A}\,.
\eeq
(Here we have introduced the correction factor $\tilde\omega>1$ to account for the deviation of the exact solution from the Keplerian law 
near $R_A$). 

As $u_R(R_A)$ is smaller than the free-fall velocity, the above formula
implies that $\omega_m< \omega_K(R_A)$, lower than 
the Keplerian angular frequency.  
For self-consistency
the coefficient $C$ in the Prandtl law is determined, according to 
\Eq{omega_m}, by the 
ratio of the radial velocity $u_R$ to the rotational velocity of matter $U_\phi$:
\beq{C1}
C=\frac{4}{9}\tilde\omega \frac{|u_R(R_A)|}{\omega_mR_A}
=\frac{4}{9}\tilde \omega \frac{|u_R(R_A)|}{U_\phi(R_A)}\,.
\eeq 
Note that this ratio is independent of 
the radius $R$ and is actually constant across the shell. Indeed, 
the radial dependence of the velocity $u_R$ 
follows from the continuity equation with account for 
the density distribution \Eq{rho_R}
\beq{u_RR}
u_R(R)=u_R(R_A)\myfrac{R_A}{R}^{1/2}\,.
\eeq
For the quasi-Keplerian law $u_\phi(R)\sim 1/R^{1/2}$, so the ratio $u_R/U_\phi$
is constant. 
Finally, 
the angular frequency of the shell rotation near the magnetosphere $\omega_m$ 
is related to the angular frequency of the motion of matter near the bow-shock as
\beq{}
\omega_m=\tilde\omega\omega_B\myfrac{R_B}{R_A}^{3/2}
\eeq
In fact, when approaching $R_A$, the integration
constant $D$ (which we neglected at large distances $R\gg R_A$) should be taken into account.
The rotational law will thus be different from the Keplerian one.

We stress the principal difference between this regime of accretion
and disc accretion. For disc accretion the radial velocity is much smaller than the
turbulent velocity, and the tangential velocity is almost Keplerian and is much larger than the
turbulent velocity.
 The radial velocity in the quasi-spherical case is not determined by
the rate of the angular momentum removal. It is determined only by the 
"permeability" of the neutron star magnetosphere for infalling matter.  
In our case we assumed it to be of the order of the convection velocity.
The tangential velocity for
the obtained quasi-Keplerian law is much smaller than the velocity of 
convective motions in the shell. 
Note also that in the case of disc accretion the turbulence can be parametrized by only one dimensionless parameter $\alpha\approx u_t^2/u_s^2$ with $0 < \alpha <1$ 
(Shakura \& Sunyaev 1973).
The matter in an accretion disc differentially rotates
with a supersonic (almost Keplerian) velocity, while in our case the shell rotates differentially
with a substantially subsonic velocity at any radius, and the turbulence in the shell is essentially subsonic. 
Of course, our case with an extended shell is strongly different from the regime of freely falling matter
with a standing shock above the magnetosphere (Arons \& Lea 1976a).

\subsection{The case without accretion}

Now let us consider the case when the plasma can not enter the magnetosphere
and no accretion onto the neutron star occurs. 
This case is similar to the subsonic propeller regime considered by Davies and Pringle (1981). \Eq{amt1} then takes the form:
\beq{amt01}
0=4\pi \rho \nu_t R^4 \frac{\partial \omega}{\partial R}+D\,.
\eeq 
(We remind the reader that the constant $D$ is determined by the spin-down
rate of the neutron star, $D=I\dot \omega^*<0$).
Solving this equation as above, we find for the rotation law in this case:
\beq{omega0}
\omega(R)=\omega_m\myfrac{R_A}{R}^{7/4}\,,
\eeq
where
\beq{}
\omega_m=\frac{I|\dot \omega^*|}{7\pi\rho(R_A)\nu_t(R_A)R_A^3}\,.
\eeq
From \Eq{nut} we find
\beq{nut1}
\nu_t(R_A)=\frac{7}{4}C\omega_mR_A^2\,,
\eeq
so for $\omega_m$ we obtain:
\beq{A45}
\omega_m=\frac{2}{7}\myfrac{I|\dot \omega^*|}{\pi C\rho(R_A)R_A^5}^{1/2}\,.
\eeq
On the other hand, $\omega_m$ is now related to the bow-shock region parameters as
\beq{}
\omega_m=\omega_B\myfrac{R_B}{R_A}^{7/4}\,.
\eeq

\section{The case with anisotropic turbulence}

The Prandl rule for viscosity used above that relates the scale and velocity of turbulent pulsations with the average angular velocity is commonly used when the turbulence is generated by the shear itself. In our problem, the turbulence is initiated by large-scale convective motions in the gravitational field. During convection, a strong anisotropic
turbulent motions can appear (the radial dispersion of chaotic motions could be much larger than the dispersion in the tangential direction), and the Prandl law could be 
inapplicable. 
Anisotropic turbulence is much more complicated and remains a poorly studied case. 

As a first step, we can adopt the empirical law for $W_{R\phi}$ suggested 
by Wasiutinski (1946):
\beq{aW1}
W_{R\phi}=2\rho(-\nu_t+\nu_r)\omega +\nu_r\rho R\frac{d\omega}{dR}
\eeq
where the radial and tangential kinematic viscosity coefficients are 
\[
\nu_r=C_\parallel\langle |u_\parallel^t|\rangle R
\] 
\[
\nu_t=C_\perp\langle |u_\perp^t|\rangle R
\] 
respectively. Dimensiionless constants $ C_\parallel$ and $C_\perp$ are of the order of one. 
In the isotropic case, $\nu_r=\nu_t$, $W_{R\phi}\sim d\omega/dR$, and in the strongly 
anisotropic case, $\nu_r\gg\nu_t$, $W_{R\phi}\sim d(\omega R^2)/dR$. 
Using these definitions, let us substitute \Eq{aW1} in \Eq{amt00} and rewrite the latter 
in the form:
\beq{}
\omega R^2\left(1-\frac{2C_\perp\langle |u_\perp^t|\rangle}{|u_R|}\right)=
C_\parallel\frac{\langle |u_\parallel^t|\rangle }{|u_R|}\frac{Rd(\omega R^2)}{dR}-\frac{D}{|\dot M|}
\eeq

We note that due to self-similarity of the shell structure $u^t_\parallel\sim 
u^t_\perp\sim u_R\sim R^{-1/2}$, so that the ratios $\langle |u_\parallel^t|\rangle/u_R$
and  $\langle |u_\perp^t|\rangle/u_R$ are constant.
In this case the obvious solution to this equation reads:
\beq{}
\omega R^2+\frac{D}{|\dot M|}\frac{1}{1-2C_\perp\frac{\langle |u_\perp^t|\rangle}{|u_R|}}=
\left[\omega_BR_B^2+\frac{D}{|\dot M|}\frac{1}{1-2C_\perp\frac{\langle |u_\perp^t|\rangle}{|u_R|}}\right]\myfrac{R_B}{R}^{\frac{|u_R|}{C_\parallel\langle |u_\parallel^t|\rangle}\left(1-2C_\perp\frac{\langle |u_\perp^t|\rangle}{|u_R|}\right)}
\eeq
(here the integration constant is defined such that $\omega(R_B)=\omega_B$).

Now let us consider the equilibrium situation where $\dot\omega^*=0$. In this case, as
we remember, 
\[
\frac{D}{|\dot M|}=-z\omega^* R_A^2\;,
\omega_m=(1-z/Z)\omega^*\,.
\]

First, let us consider the case of strongly anisotropic, almost radial turbulence where $\langle |u_\perp^t|\rangle=0$. In this case, the specific angular momentum at the Alfven radius is 
\beq{case1}
\omega_m R_A^2\left[1+\frac{z}{1-z/Z}\left(\myfrac{R_B}{R_A}^
\frac{|u_R|}{C_\parallel\langle|u^t_\parallel|\rangle}-1\right)\right]=
\omega_BR_B^2\myfrac{R_B}{R_A}^\frac{|u_R|}{C_\parallel\langle|u^t_\parallel|\rangle}\,.
\eeq
It is seen that in the case of very weak accretion (or, in the limit, when the accretion through magnetosphere ceases at all), $|u_R|\ll C_\parallel\langle|u^t_\parallel|\rangle$, an almost iso-angular-momentum
rotation in the shell is realized.

The next case is where the anisotropy is such that 
$C_\perp\langle|u^t_\perp|\rangle/|u_R|=1/2$. 
Then we have strictly iso-angular-momentum distribution: $\omega_mR_A^2=\omega_BR_B^2$. 

If the turbulence is fully isotropic: $C_\perp\langle|u^t_\perp|\rangle=\
C_\parallel\langle|u^t_\parallel|\rangle=\tilde C\langle|u^t|\rangle$. Denoting $\epsilon=|u_R|/(\tilde C\langle |u^t|\rangle)$ 
we find:
\beq{case2}
\omega_mR_A^2\left[1+\myfrac{z}{1-z/Z}\myfrac{1}{2/\epsilon-1}
\left(1-\myfrac{R_A}{R_B}^{2-\epsilon}\right)
\right]=\omega_BR_B^2\myfrac{R_A}{R_B}^{2-\epsilon}
\eeq
Note that if $\epsilon\to 0$ (there is no accretion thorugh the magnetosphere), $\omega_m\to \omega_B$, 
i.e. a solid-body rotation without accretion is established (cf. the case one above!).
In the case for $\epsilon=3/2$, a near quasi-Keplerian angular rotation
law can be realized. We renmind that a similar quasi-Keplerian law was obtained in 
Appendix A above with the use of the Prandtl rule for isotropic turbulent viscosity. 
This was the only solution in that case. Here, in contrast, the quasi-Keplerian law
is only a partucular case of the general solution obtained with the 
use of the Wasiutinsky rule for
anisotropic turbulent viscosity.

As we have shown in the main text, the case of quasi-Kpeleerian rotation law 
is not favored by observations. So we conclude that most likely near iso-angular-momentum 
rotational velocity distribution with anisotropic turbulence initiated by convection in the shell is realized. We remind that in thin accretion discs where the vertical height 
limits the scale of turbulence, the Prandlt law for viscosity works very well
(Shakura \& Sunyaev, 1973). 

\section{Corrections to the radial temperature gradient}

Here we shall estimate how the radial temperature 
gradient differs from the adiabiatic law due
to convective motions in the shell.
By multiplying \Eq{sd_om} by $(1/2)(\omega_m-\omega^*)$, we obtain the
convective heating rate 
caused by interaction of the shell with the magnetosphere:
\beq{}
L_c=\frac{1}{2}Z\dot M R_A^2(\omega_m-\omega^*)^2\,.
\eeq
Multiplying the same \Eq{sd_om} by $\omega^*$ yields the rate of change of the mechanical energy of the neutron star 
\beq{}
L_k=Z\dot M R_A^2\omega^*(\omega_m-\omega^*)\,.
\eeq
The total energy balance is then
\beq{}
L_t=L_c+L_k=\frac{1}{2}Z \dot M R_A^2 (\omega_m^2-\omega^{*2})\,.
\eeq
Note that the obtained formula for $L_c$ is similar to that describing
energy release in the boundary layer of an accretion disc, see Shakura \& Sunyaev (1988). 

The convective energy flux is:
\beq{q_m}
q_c=\frac{L_c}{4\pi R^2}=\frac{Z\dot M R_A^2(\omega_m-\omega^*)^2}{8\pi R^2}\,.
\eeq

On the other hand, the convective energy flux can be related to the entropy gradient (see \citet{ssz})
\beq{8}
q_c=-\rho\nu_c T\frac{dS}{dR}\,,
\eeq
where $S$ is the specific entropy (per gram). Here $\nu_c$ 
is the radial turbulent heat conductivity 
\beq{}
\nu_c=<u_c l_c>=C_hu_cR
\eeq
where $l_c\sim R$, the convective velocity $u_c\sim c_s\sim R^{-1/2}$ and $C_h$ is a numerical
coefficient of the order of one. So 
\beq{}
\nu_c=\nu_c(R_A)\myfrac{R}{R_A}^{1/2}\,.
\eeq

Next, we make use of the thermodynamic identity for the specific enthalpy $H$
\beq{tdid}
\frac{dH}{dR}=\frac{1}{\rho}\frac{dP_g}{dR}+T\frac{dS}{dR}\,.
\eeq
We remind the reader that the enthalpy can be written as
\[
dH=c_p dT\,.
\]
where
\[
c_p=T\myfrac{\partial S}{\partial T}_p=\frac{\gamma}{\gamma-1}\frac{{\cal R}}{\mu_m}
\]
is the specific heat capacity at constant pressure. 
Expressing $T(dS/dR)$ from 
\Eq{8} and making use of the hydrostatic equation [\Eq{hse_sol}] written as
\[
\frac{dP_g/\rho}{dR}=-\frac{{\cal R}}{\mu_m c_p}\frac{GM}{R^2}\psi(\gamma,m_t)\,,
\]
the thermodynamic identity \Eq{tdid} can be rewritten in the form
\beq{tdid1}
\frac{dT}{dR}=-\frac{1}{c_p}\left[\frac{GM}{R^2}\psi(\gamma,m_t)
-\frac{Zu_R(R_A)}{2\nu_c(R_A)}\myfrac{R_A}{R}R_A^2(\omega_m-\omega^*)^2\right]\,.
\eeq
By definition the adiabatic temperature gradient is determined by the first term
on the right side, $(dT/dR)_{ad} =g/c_p$. 
Equation (\ref{tdid1}) can be integrated to find the actual
temperature dependence:
\beq{Treal}
T=\frac{1}{c_p}\left[
\frac{GM}{R}\psi(\gamma,m_t)
-\frac{Zu_R(R_A)}{2\nu_c(R_A)}R_A^3(\omega_m-\omega^*)^2\ln\myfrac{R}{R_A}
\right]\,.
\eeq

Close to the equilibrium ($I\dot \omega^*=0$), we can use \Eq{do} and write
\beq{}
T=\frac{1}{c_p}\left[\frac{GM}{R}\psi(\gamma,m_t)
-\frac{u_R(R_A)}{2C_h u_c(R_A)}\omega^{*2}R_A^2\frac{z^2}{Z}\ln\myfrac{R}{R_A}\right]\,.
\eeq

This solution shows that in the whole region between $R_A$ 
and $R_B$, for slowly rotating pulsars (i.e.,
in which $\omega_m\ll\omega_K(R_A)$), the temperature distribution is close to the adiabatic law with a
temperature gradient close to the adiabatic one [\Eq{hse_sol}]:
\beq{}
T\approx \frac{\gamma-1}{\gamma}\frac{GM}{{\cal R}R}\psi(\gamma,m_t)\,.
\eeq

In fact, we have only taken into account the energy release due to the frequency difference 
near the magnetosphere. In fact, there can be additional sources of energy in the shell (e.g. the heat release during magnetic reconnection and turbulence (see Appendix D), etc.). 

\section{Dynamics of a stationary spherically-symmetric ideal gas flow}

In this Appendix, we write down the gas-dynamic equations of the 
spherically symmetric ideal gas flow onto a Newtonian gravitating center.
This problem was considered in the classical paper by Bondi (1952) for an adibatic accretion 
flow. Adiabatic gas 
outflows (stellar winds) were studied by Parker (1963). We focus on 
the role of the cooling/heating processes near the Alvenic surface.
As we discussed in the main text, at low X-ray luminosities 
the quasi-static shell is capable of removing the angular momentum from the rotating
magnetosphere via convective motions. As the accretion rate exceeds some critical value, 
strong Compton cooling causes a free-fall gap to appear above the magnetosphere,
and the angular momentum cannot be transferred from the magnetosphere to the shell any more. 
     
The equation of motion \Eq{v_R1} in the absence of viscosity reads:
\beq{eom}
u\frac{du}{dR}=-\frac{1}{\rho}\frac{dP_g}{dR}-
\frac{1}{\rho}\frac{dP_\parallel^t}{dR}-\frac{2(P_\parallel^t-P_\perp^t)}{\rho R}-\frac{GM}{R^2}
\eeq
Here $P_g=\rho c_s^2/\gamma$ is the gas pressure, and $P^t$ stands for the pressure due to turbulent pulsations, which in general are anisotropic:
\beq{}
P_\parallel^t =\rho <u_\parallel^2>=\rho m_\parallel^2 c_s^2=\gamma P_g  m_\parallel^2
\eeq
\beq{}
P_\perp^t =2\rho <u_\perp^2>=2\rho m_\perp^2 c_s^2 =2\gamma P_g  m_\perp^2
\eeq
($<u_t^2>=<u_\parallel^2>+2<u_\perp^2>$ is the  
turbulent velocity dispersion, $m_\parallel^2$ and $m_\perp^2$ are the parallel and 
perpendicular turbulent Mach numbers squared). 

From the first law of thermodynamics 
\beq{1td}
\frac{dE}{dR}=\frac{P_g}{\rho}\frac{d\rho}{dR}+T\frac{dS}{dR}
\eeq
where the specific internal energy (per gram) is 
\beq{E}
E=c_VT=\frac{c_s^2}{\gamma(\gamma-1)}\,,
\eeq
the heat capacity is 
\beq{cV}
c_V=\frac{{\cal R}}{\mu_m}\frac{1}{\gamma-1}
\eeq
From the second law of thermodynamics, 
the specific entropy change can be written through the 
specific heat change rate $dQ/dt$ [erg/s/g] as 
\beq{}
T\frac{dS}{dR}=\frac{dQ}{dR}=\frac{dQ/dt}{u}\,.
\eeq
Using the mass continuity equation
\beq{dotM}
\dot M=4\pi R^2\rho u
\eeq
we find 
\beq{}
\frac{1}{\rho}\frac{d\rho}{dR}=-\frac{2}{R}-\frac{1}{2u^2}\frac{du^2}{dR}\,.
\eeq
Using the relation $c_s^2=\gamma{\cal R}T$, we finally obtain:
\beq{dcdR}
\frac{1}{c_s^2}\frac{dc_s^2}{dR}=(\gamma-1)\left[-\frac{2}{R}-\frac{1}{2u^2}\frac{du^2}{dR}\right]+
\frac{dQ/dt}{uc_VT}\,.
\eeq
Note that this equation can also be derived 
directly from the ideal gas equation of state written 
in the form
\beq{eos}
P_g=Ke^{S/c_V}\rho^\gamma
\eeq
where $K$ is a constant.

Using \Eq{dcdR}, the gas pressure gradient
can be rewritten in the form:
\beq{dPdR2}
\frac{1}{P_g}\frac{dP_g}{dR}=\frac{c_s^2}{c_V u}\frac{dQ/dt}{T}
+c_s^2\left[-\frac{2}{R}-\frac{1}{2u^2}\frac{du^2}{dR}\right]
\eeq 
Plugging \Eq{dPdR2} into the equation of motion finally yields:
\beq{eom1}
\frac{1}{2}\frac{1}{u^2}\frac{du^2}{dR}=
\left[c_s^2(1+\gamma m_\parallel^2)
\left(\frac{2}{R}-\frac{dQ/dt}{c_V uT}\right)
-2c_s^2\frac{(m_\parallel^2-m_\perp^2)}{R}-\frac{GM}{R^2}
\right]/\left[u^{2^{}}-c_s^2(1+\gamma m_\parallel^2)\right]\,.
\eeq 
Note also that in the strongly anisotropic case where $m_\parallel^2=m_t^2\gg m_\perp^2$, the role of 
turbulence increases in comparison with the isotropic case where $m_\parallel^2=m_\perp^2=(1/3) m_t^2$.

We can also introduce the Mach number in the flow ${\cal M}=u/c_s$. Then from \Eq{dcdR} and
\Eq{eom1} we derive the equation for the Mach number:
\begin{eqnarray}
\label{mach}
&\frac{[{\cal M}^2-(1+\gamma m_\parallel^2)]}{{\cal M}^2}
\frac{d{\cal M}^2}{dR}=\nonumber\\
& \left\{
\frac{2\left[(\gamma-1){\cal M}^2-(\gamma+1)(m_\parallel^2-m_\perp^2)\right]}{R}-
\frac{\left[{\cal M}^2+\gamma(1+\gamma m_\parallel^2)\right]}{c_VT}\frac{dQ}{dR}-
\frac{(\gamma+1)GM}{R^2c_s^2}
\right\}\,.
\end{eqnarray}
where we have substituted $(dQ/dt)=u(dQ/dR)$. Equations (\ref{dcdR}), (\ref{eom1}) and
(\ref{mach}) can be used to solve the dynamics of the accretion flow for pairs of independent variables $(u, c_s)$, $(u,{\cal M})$ or $(c_s, {\cal M})$.  Here, however, we shall
only consider the behaviour of the flux near the singular point. To this end, we can use \Eq{eom1}.

\Eq{eom1} has a singular saddle point where the denominator vanishes: 
\beq{saddle}
u^2=c_s^2(1+\gamma m_\parallel^2)
\eeq 
So must the numerator, from which we find the quadratic equation for the velocity at the singular point:
\beq{qe}
u^2\frac{2}{R}\myfrac{1+(\gamma-1)m_\parallel^2+m_\perp^2}{1+\gamma m_\parallel^2}
-u\myfrac{dQ/dt}{c_V T}-\frac{GM}{R^2}=0\,.
\eeq
Remember that in 
the adiabatic case ($dQ/dt=0$) without turbulence 
at the saddle point we have simply 
\beq{cs2ad}
u^2=c_s^2=\frac{GM}{2R}\,.
\eeq 
We stress that the presence of turbulence increases the velocity at the singular point. For example, for $\gamma=5/3$ we find for strong anisotropic turbulence $u^2=c_s^2(1+(5/3)m_\parallel^2)$; for the isotropic turbulence the correction is smaller:  $u^2=c_s^2(1+(5/9)m_t^2)$. 
The transition through the sound speed (the sonic point where $u^2=c_s^2$) lies above the saddle point due to turbulence, and there is no singularity in the sonic point. 

First let us determine the turbulence heating rate in the quasi-static shell 
$(dQ/dt)^+_t$: 
\beq{}
\myfrac{dQ}{dt}^+_t=\frac{1}{2}\frac{<u_t^2>}{t_t}
\eeq
where the characteristic time of the turbulent heating is 
\beq{}
t_t=\alpha_t\frac{R}{u_t}=\alpha_t\frac{R}{m_t c_s}
\eeq
with $\alpha_t$ being a dimensionless constant characterizing the turbulent dissipation 
energy rate 
and the turbulent Mach number is $m_t^2\equiv m_\parallel^2+2m_\perp^2$.
The turbulent heating rate can thus be written as
\beq{}
\myfrac{dQ}{dt}^+_t=\frac{c_s^3}{2\alpha_t R}m_t^3\,.
\eeq

In the case of Compton cooling we have
\beq{}
\myfrac{dQ}{dt}_C^-=-\frac{c_V (T-T_x)}{t_C}
\eeq
where $t_C$ is the Compton cooling time [\Eq{t_comp}]. 

\Eq{qe} can now be written in the form:
\beq{qe1}
u^2\frac{2}{R}\myfrac{1+(\gamma-1)m_\parallel^2+m_\perp^2}{1+\gamma m_\parallel^2}
-u^2\frac{c_s}{u}\frac{\gamma(\gamma-1)m_t^3}{2\alpha_t R}
+\frac{u(1-T_x/T)}{\gamma t_C}-\frac{GM}{R^2}=0\,.
\eeq
As our problem is one of accretion, the sign of the velocity $u=dR/dt$ is negative,
so below we shall write $u=-|u|$. Then for the absolute value of the velocity at the
singular point where $c_s/|u|=-1/(1+\gamma m_\parallel^2)^{1/2}$ 
we have the quadratic equation:
\beq{qe2}
u^2\frac{2}{R}\myfrac{1+(\gamma-1)m_\parallel^2+m_\perp^2}{1+\gamma m_\parallel^2}
+u^2\frac{1}{(1+\gamma m_\parallel^2)^{1/2}}\frac{\gamma (\gamma-1)m_t^3}{2\alpha_t R}
-\frac{|u|(1-T_x/T)}{\gamma t_C}-\frac{GM}{R^2}=0\,.
\eeq
 
In this case, the solution  to \Eq{qe} reads:
\beq{u_sol}
|u|=\frac{R(1-T_x/T)}{4\gamma t_C A}+\sqrt{\frac{2GM}{R}}
\left[\frac{1}{4A}+\frac{R}{2GM}\frac{R^2(1-T_x/T)^2}{16\gamma^2t_C^2A^2}\right]^{1/2}
\eeq
where we have introduced the dimensionless factor 
\beq{A}
A=\frac{1+(\gamma-1)m_\parallel^2+m_\perp^2}{1+\gamma m_\parallel^2}+
\frac{\gamma (\gamma-1)( m_\parallel^2+2m_\perp^2)^{3/2}}{4\alpha_t (1+\gamma m_\parallel^2)^{1/2}}\,.
\eeq
In the case of isotropic turbulence where $m_\parallel=m_\perp=1/\sqrt{3}, m_t=1$, for $\gamma=5/3$ 
the factor $A\approx 1.23$, and in the case of strongly anisotropic turbulence where
$m_\parallel=1, m_\perp=0, m_t=1$, this factor is $A\approx 0.8$.

In units of the free-fall velocity the solution \Eq{u_sol} reads:
\beq{cs1}
f(u)=\frac{|u|}{u_{ff}}=\frac{(1-T_x/T)}{4\gamma A}\myfrac{t_{ff}}{t_C}+
\frac{1}{2}
\left[\frac{1}{A}+\frac{(1-T_x/T)^2}{4\gamma^2A^2}\myfrac{t_{ff}}{t_C}^2\right]^{1/2}\,.
\eeq

With Compton cooling, the temperature changes exponentially:
\beq{}
T=T_x+(T_{cr}-T_x)e^{-t/t_C}
\eeq
(see the main text).  
When cooling is slow, $t_{ff}/t_C\ll 1$, 
the critical point lies inside the Alfven surface, i.e. no transition through the
critical point occurs in the flow before it meets the magnetosphere, and in this case 
we expect that the accretion settling regime can be realized. 
If this point lies abobe the Alfven surface, the velocity of the flow
can become supersonic above the magnetosphere, and one can expect the
formation of a shock. Both turbulence and rapid cooling shifts
the location of the critical point upward in the flow. 
 
In the case of rapid cooling, $t_{ff}/t_C\gg 1$, 
$T\to T_x$, so again $u/u_{ff}\approx 1/2$, but the critical point rises  
above the Alfven surface, so a free-fall gap above the magnetosphere 
appears. The ratio $f(u)=|u|/u_{ff}$ reaches a maximum at  $t_{ff}/t_C\approx 0.46$ 
(assuming a typical ratio $T_{cr}/T_x=10$), and depending on the value of $A=0.8\div 1.23$ (anisotropic or isotropic turbulence) 
it equals $f(u)=0.5-0.6$.
 
\section*{Acknowledgments}
We are very grateful to the anonymous referee for careful reading of
the manuscript and many useful comments. 
The authors also acknowledge  Dr. 
V. Suleimanov (IAAT) for  
discussions and valuable notes, Dr. V. Doroshenko (IAAT) for courtesly providing
the torque-luminosity plots for GX 301-2 and Vela X-1, and Ms. A. Gonz\'alez-Gal\'an
for providing data on GX 1+4 prior to publication. 
NIS thanks the Max-Planck Institute for Astrophysics (Garching) for the hospitality. The work by NIS, KAP and AYK is supported by RFBR grants 09-02-00032, 12-02-00186, and 10-02-00599. LH is supported
by a grant from the Wenner-Gren foundations.

\end{document}